\documentclass[sigconf]{acmart}

\acmConference[arXiv '20]{arXiv '20}{Jan 30, 2020}
 
\usepackage[toc,page]{appendix}
\usepackage{graphicx}			 
\usepackage{amssymb}
\usepackage{amsmath}
\usepackage{caption}
\usepackage{url}
\usepackage{amssymb}
\usepackage{amsthm}
\usepackage{subcaption}
\usepackage{xcolor}
\usepackage{enumitem}
\usepackage{multicol}
\makeatletter
\def\plist@algorithm{Algorithm\space}
\makeatother
\usepackage[noend]{algpseudocode}
\usepackage{algorithm}
\usepackage{multirow}
\usepackage{pgfplots}
\usepackage{tikz}

\newcommand{\ms}[1]{%
  \relax\ifmmode
  \mathord{\mathcode\setminus-=``702D\it #1\mathcode\setminus-=''2200}%
  \else
  {#1}%
  \fi
}

\begin{document}
\title{Privacy-Preserving Search for a Similar Genomic Makeup in the Cloud}

\author{Xiaojie Zhu}
\email{xiaojiez@ifi.uio.no}
\affiliation{University of Oslo}

\author{Erman Ayday}
\email{erman208@case.edu}
\affiliation{Case Western Reserve University}

\author{Roman Vitenberg}
\email{romanvi@ifi.uio.no}
\affiliation{University of Oslo}

\author{Narasimha Raghavan Veeraragavan}
\email{vnragavan@protonmail.com}
\affiliation{University of Oslo}



\begin{abstract}
Increasing affordability of genome sequencing and, as a consequence, widespread availability of genomic data opens up new opportunities for the field of medicine, as also evident from the emergence of popular cloud-based offerings in this area, such as Google Genomics. To utilize this data more efficiently, it is crucial that different entities share their data with each other. However, such data sharing is risky mainly due to privacy concerns. In this paper, we attempt to provide a privacy-preserving and efficient solution for the ``similar patient search'' problem among several parties (e.g., hospitals) by addressing the shortcomings of previous attempts. We consider a scenario in which each hospital has its own genomic dataset and the goal of a physician (or researcher) is to search for a patient similar to a given one (based on a genomic makeup) among all the hospitals in the system. To enable this search, we let each hospital encrypt its dataset with its own key and outsource the storage of its dataset to a public cloud. The physician can get an authorization from multiple hospitals and send a query to the cloud, which efficiently performs the search across authorized hospitals using a privacy-preserving index structure. We propose a hierarchical index structure to index each hospital's dataset with low memory requirement. Furthermore, we develop a novel privacy-preserving index merging mechanism that generates a common search index from individual indices of each hospital to significantly improve the search efficiency. We also consider the storage of medical information associated with genomic data of a patient (e.g., diagnosis and treatment). We allow access to this information via a fine-grained access control policy that we develop through the combination of standard symmetric encryption and ciphertext policy attribute-based encryption. Using this mechanism, a physician can search for similar patients and obtain medical information about the matching records if the access policy holds. We conduct experiments on large-scale genomic data and show the efficiency of the proposed scheme. Notably, we show that under our experimental settings, for large query sizes (e.g., when the query includes a large portion of a patient's genotype), 
the proposed scheme is more than $60$ times faster than Wang et al.'s protocol~\cite{wang2015efficient} and more than $95$ times faster than Asharov et al.'s~\cite{asharov2017privacy} and Thomas et al.'s~\cite{schneider2019episode} solutions. 
\end{abstract}




\maketitle
\vspace{-10pt}
\section{Introduction}

Thanks to the sharp cost reduction in the whole genome sequencing, today, digital genomes are used in many applications such as paternity tests~\cite{marshall1998statistical}, personalized medicine~\cite{weston2004systems}, and genetic compatibility tests~\cite{gusella1983polymorphic}. Among these uses of genomic data, arguably the most important one is in healthcare. Physicians now treat their patients based on their genetic makeup. They provide different prescriptions to patients having the same disease but with different genetic makeup. Thus, it is very valuable for a physician to identify other patients that are in similar conditions to their patient to get more insight about the diagnosis and treatment procedures. However, doing such a search in a broad fashion (e.g., nation-wide) has many unique challenges as we discuss in the following.

\noindent\textbf{Confidentiality of genomic data and affiliated sensitive information (ASI).} There has already been several privacy concerns raised related to genomic data~\cite{gymrek2013identifying, shringarpure2015privacy}. Since genomic data includes information about an individual's phenotype, ethnicity, family members, disease conditions, and more, if it falls into wrong hands, the consequences may be as serious as genetic discrimination (e.g., in healthcare or employment). Genomic data is often associated with the medical condition of a patient, including diagnosis, treatment, and symptoms. We refer to this information as ``affiliated sensitive information'' (ASI) of the patient. For instance, a mutation in the BRCA gene is recognized as a major contributor for breast cancer~\cite{cowper2012breast}, similarly the treatment of many cancer types are determined based on the genetic makeup of a patient. Protecting the confidentiality of patients' genomic data and ASI is essential for the hospitals, and hence, hospitals are not willing to open their datasets to each other or share their datasets with a public cloud service provider (CSP). Therefore, genomic data sharing mechanisms that provide privacy guarantees to the hospitals about their datasets are required to pave the way to an efficient and privacy-preserving nation-wide similar patient search protocol.

\noindent\textbf{Efficiency of the search process.} The search process to identify a target genome sequence (i.e., a similar patient) should be efficient. However, considering the scale of genomic data and the scale of the search (i.e., number of hospitals and the number of patients in each hospital), providing an efficient protocol along with the privacy goals is not trivial. One obvious approach is to apply index structures (e.g., suffix tree, prefix tree, or binary tree) to make the search process more efficient. However, such indexing techniques cannot be directly applied for the genome search due to (i) size of genomic data, (ii) variation of genomic data between individuals (e.g., mutations), and (iii) the aforementioned privacy requirements. Thus, new techniques are required to provide both privacy guarantees and efficiency for similar patient search problem.    

\noindent\textbf{Search over several parties.} Searching for similar patients is more effective and helpful if the physician can search datasets of more hospitals. Previous studies assume the physician to query all hospitals individually (one-by-one), however such a strategy is both time consuming and unreliable since it requires the cooperation of each hospital in real-time. Instead, it would be easier if all hospitals outsource their datasets to a common entity (e.g., a CSP) and the physician directly queries this CSP. However, such an approach is not trivial due to privacy concerns. Furthermore, as new hospitals join the system, or as the datasets of the existing hospitals change, data stored at the CSP should be updated and this may cause additional cost. Therefore, we need new solutions to share datasets among several hospitals in a privacy-preserving and efficient way.


In this paper, to the best of our knowledge, we propose the first framework to tackle all these challenges. We propose a scheme in which each hospital encrypts its own dataset (with its unique key) and outsources the storage and processing for the search operation to a CSP. For privacy, we encrypt genomic data with a standard encryption algorithm and propose a novel indexing mechanism for privacy-preserving search. 
The proposed indexing mechanism provides not only privacy, but also the ability to search over several hospitals' datasets in an efficient way. Each hospital encrypts its own dataset independently while the searchability of ciphertext is enabled across all the hospitals through this indexing mechanism. 

In order to achieve efficient search and outsource computation-intensive tasks to the CSP, we propose two mechanisms to advance the proposed indexing scheme.
First, we propose a hierarchical clustering algorithm and a hierarchical index structure to accelerate the search process.
Second, we introduce a privacy-preserving index merging algorithm to avoid CSP sequentially searching over all the stored hierarchical index structures (e.g., belonging to different hospitals) one-by-one. 
To enable the ASI to be properly accessible by legitimate clients, we also introduce an ASI sharing scheme. Considering the fine-grained access requirement, we adopt chosen policy attribute-based encryption (CPABE). 
In addition, to enable participants to use different secret keys to encrypt the ASI, we introduce a re-encryption mechanism. 
We implement and evaluate the proposed scheme under various scenarios. 
Also, we show that compared with the state-of-the-art, the proposed scheme performs more than $60$ times faster than Wang et al.'s protocol~\cite{wang2015efficient} and more than $95$ times faster than Asharov et al.'s~\cite{asharov2017privacy} and Thomas et al.'s~\cite{schneider2019episode} schemes, especially for large query sizes. 


\vspace{-5pt}
\section{Related work}\label{sec:relatedwork}
\vspace{-5pt}

Privacy of genomic data has been recently a very active research topic~\cite{naveed2015privacy}. Several privacy-preserving solutions have been proposed for processing of genomic data in different settings, including personalized medicine~\cite{baldi2011countering}, research~\cite{kantarcioglu2008cryptographic,lu2015efficient}, alignment~\cite{chen2012large}, and management of raw genomic data~\cite{ayday2013privacy}. 

There has been many earlier work on privacy-preserving pairwise comparison of genomes (or identification of a pattern in a given DNA sequence). Atallah et al. proposed a privacy-preserving edit distance protocol based on dynamic programming~\cite{atallah2005secure}. Computational efficiency of this work was later improved by Jha et al.~\cite{jha2008towards}. Troncoso-Pastoriza et al. proposed a protocol to execute finite state machine (FSM) in an oblivious manner~\cite{troncoso2007privacy}. 
Yasuda et al. applied somewhat homomorphic encryption (SWHE) to implement privacy-preserving Hamming distance computation of two genome sequences~\cite{yasuda2013secure}. Cheon et al. used SWHE to implement secure edit distance computation of two genome sequences~\cite{cheon2015homomorphic}. Wang et al. proposed a scheme for DNA sequence matching with only one-round of interaction~\cite{wang2017privacy}. 
Sousa et al. combined SWHE and private information retrieval (PIR) to implement secure search over outsourced VCF files~\cite{sousa2017efficient}. Cheng et al. proposed secret sharing (using two non-colluding public clouds) for similarity computation between genome sequences~\cite{cheng2018secure}. Although these schemes are useful for pairwise comparison of genomes (or comparison of a pattern and a genome), they cannot be generalized for 1-to-n comparison between the genomes easily due to efficiency and practicality issues, and hence they are not applicable for the similar patient search problem.

Similar to our proposed work, privacy-preserving similar patient search has been considered by a few works. Wang et al. proposed an efficient genome-wide, privacy-preserving similar patient query scheme for two parties~\cite{wang2015efficient}. In their scheme, the edit distance of two genome sequences is transformed into finding the number of different elements between two sets. 
Asharov et al. addressed the same problem by 
pre-processing genome sequences into proper fragments before comparison. 
Both these works assume that genomic data is stored at local datasets (e.g., each hospital storing its own genomic dataset) and the client (physician) looks for the top k-closest sequences at each local dataset. 
This makes the search process impractical since both schemes require each hospital to be available all the time and responsive to the queries. 
Schneider et al.~\cite{schneider2019episode} adapted Asharow et al.'s solution to support outsourcing. 
In a nutshell, their scheme is a secret sharing-based mechanism, in which the data owners outsource the database storage to two semi-trusted service providers. A client's query is generated by interacting with these service providers. 
Schneider et al.'s scheme relies on the existence of more than one (semi) trusted and non-colluding entities. Furthermore, hospitals are typically reluctant to outsource their medical datasets to cloud-based service providers without encryption. 
Therefore, secret sharing-based solutions, although efficient, are not practical for real-life implementation of this scenario.


\textbf{Our contribution.} As opposed to previous work, here, we provide a significantly more practical and efficient solution by letting the hospitals outsource the storage of their datasets to a cloud service provider (CSP) in a privacy-preserving way. 
To provide the privacy of outsourced data, we let each data owner (hospital) encrypt its data with its unique cryptographic key. Thus, as opposed to similar work that use a CSP to process data from multiple sources, we avoid single point-of-failure by encrypting all the outsourced data with different keys. 
We also consider a dynamic system in which new hospitals join by uploading their datasets to the CSP in an efficient way. 
We provide these functionalities via a novel indexing scheme and a novel privacy-preserving index merging algorithm. 
Our evaluation results on real genomic data shows that the proposed scheme provides more than $60$ times better performance (in terms of run-time) than Wang et al.'scheme \cite{wang2015efficient} and more than $95$ times better performance than Asharov et al.'s~\cite{asharov2017privacy} and Thomas et al.'s~\cite{schneider2019episode} schemes, especially for large query sizes (e.g., including large number of point mutations).
 
Furthermore, we consider controlled access to affiliated sensitive information (ASI) such as diagnosis, treatment, or symptoms that can be associated with genomic information.
We provide fine-grained access control to ASI so that an authorized physician can not only identify similar patients but she can also obtain medical information about them. 

\vspace{-5pt}
\section{Background}
\label{sec:background}\vspace{-5pt}
Here, we provide brief background information about genomics and less-common cryptographic primitives we use in this work. We provide the background about more common primitives such as asymmetric bilinear groups and Bloom filters in Appendix~\ref{app:background}.\vspace{-5pt}

\subsection{Genomics Background}



The most common mutation in human population is called single nucleotide polymorphism (SNP). It is the variation in a single nucleotide at a particular position of the genome~\cite{risch2000searching}. There are about 5 million SNPs observed per individual and sensitive information about individuals (such as disease predispositions) are typically inferred by analyzing the SNPs. 
Two kinds of nucleotides (or alleles) are observed for each SNP: (i) major allele is the one that is observed with a high frequency and (ii) minor allele is the one that is observed with low frequency. The frequency of the minor allele in a given population is denoted as the minor allele frequency (MAF). Each SNP includes two nucleotides, one inherited from the father and the other one from mother. For simplicity, we represent the value of a SNP $i$ as the number of its minor alleles, and hence $SNP_i \in \{0, 1, 2\}$. A SNP is represented by an (ID, value) pair, where the ID is taken from a large standardized set of strings and the value is in $\{0, 1, 2\}$. In the following sections, if we mention a SNP (or SNPs) without mentioning the ID or value, we mean both parts.\vspace{-5pt}
 
\subsection{Ciphertext Policy Attribute-Based Encryption (CPABE)}
CPABE enables controlled access to encrypted data~\cite{bethencourt2007ciphertext}. It consists of the following four algorithms. 

\noindent\textbf{Setup}. Outputs the public parameters \textit{PK} and a master key $\textit{MK}_i$ given a security parameter. 
 
\noindent\textbf{Encrypt}(\textit{PK}, \textit{M}, $\mathbb{A}$). Takes as input public parameters \textit{PK}, a message $M$, and an access structure $\mathbb{A}$ over the universe of attributes. The algorithm encrypts $M$ and produces a ciphertext \textit{CT} such that only a client that possesses a set of attributes that satisfy the access structure $\mathbb{A}$ can decrypt \textit{CT}. 
 
\noindent\textbf{Key Generation}($\textit{MK}_i$, \textit{S}). Outputs a private key \textit{sk} given the master key $\textit{MK}_i$ and a set of attributes \textit{S}. 
 
\noindent\textbf{Decrypt}(\textit{PK}, \textit{CT}, \textit{sk}). Takes as input public parameters \textit{PK}, a ciphertext \textit{CT}, which contains an access policy $\mathbb{A}$, and \textit{sk}, which is a private key for a set \textit{S} of attributes. If \textit{S} satisfies the access structure $\mathbb{A}$, then the algorithm decrypts the ciphertxt \textit{CT} and returns a message \textit{M}. \vspace{-5pt}

\subsection{Customized Bloom Filter}\label{sec:bloom_filter}

Compared with the standard Bloom filter (\textit{BF}), the \emph{customized Bloom filter} (\textit{CBF}) uses one perfect hash function instead of $k$ normal hash functions. The perfect hash function~\cite{fredman1982storing} for a set of data items is a hash function that maps distinct elements in the set to a set of integers with no collisions. These integers are further utilized as indices of a bit array and corresponding values are set to $1$. In the remaining of the paper, if we do not specify the type of the Bloom filter, then it is the standard one. 

\vspace{-5pt}
\section{Problem Formulation}
\label{sec:poblemformualtion}\vspace{-5pt}
Here, we introduce our system, threat, and query models. \vspace{-5pt}

\subsection{System Model}
\begin{figure}
\includegraphics[width=0.47\textwidth]{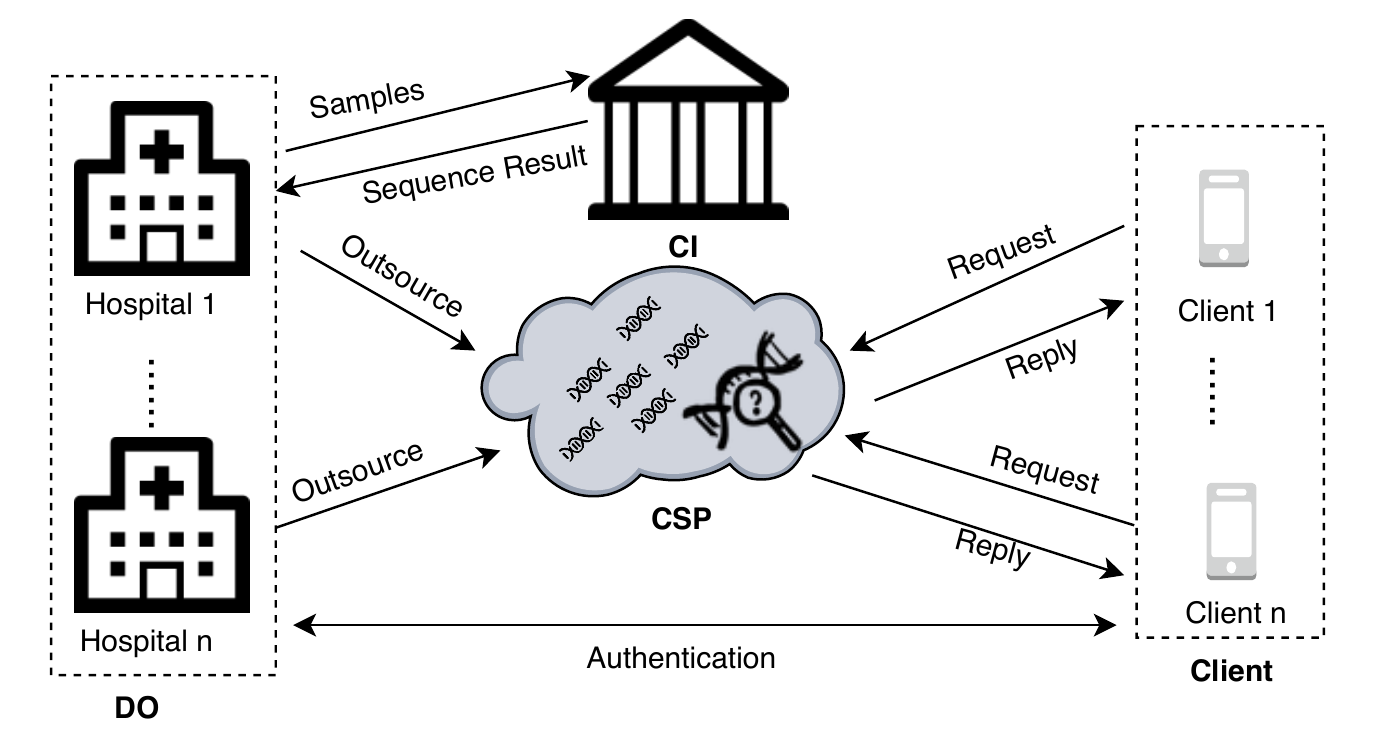}
\caption{System model. The certificated institution (CI) is responsible for sequencing the submitted samples. The data owner (DO) (e.g., hospital) processes its data and outsources it to the cloud service provider (CSP). 
If a client wants to access the outsourced data, she needs to get authorization from the DO(s). 
After obtaining the authorization from the DO, the client sends a request to the CSP for target data. The CSP processes the request and sends the result to the client.}
\vspace{-15pt}
\label{fig:sm}
\end{figure}

As shown in Figure~\ref{fig:sm}, our proposed model consists of four entities: data owner (DO), certificated institution (CI), cloud service provider (CSP), and client (e.g., physician). The DO can be considered as the hospital. The hospital collects biological samples from patients with their consent and sends the samples to the CI for sequencing. The CI is an authority or trusted institution that is responsible for sequencing DNA and generating the VCF files (the file format to store the SNPs of individuals). 
Upon receiving VCF files from the CI, hospital first processes them (e.g., generating the complementing ASI, indexing, and encryption) and then, outsources the storage of the encrypted dataset to the CSP. The CSP stores the uploaded encrypted datasets and responds to the queries of the clients for similar patient search. After a client is authenticated by a hospital (e.g., to make sure that she is a legitimate physician), she can issue a query to the CSP to search over the stored data belonging to the corresponding hospital(s). Upon receiving the search result from the CSP, the client further processes the retrieved result and obtains the plaintext response. \vspace{-5pt}

\subsection{Threat Model}\label{ssec:tm}
We assume that the CI is a trusted party, which is consistent with the previous work~\cite{ayday2013privacy, ayday2013protecting}.

The CI is only responsible for the sequencing. Due to the nature of today's sequencing technology, existence of such a trusted CI is a mandatory assumption for all existing schemes.
The CSP is assumed to be semi-honest, following the common practice in this area~\cite{wang2015efficient, asharov2017privacy}. 
Under this assumption, the CSP follows the protocol honestly and may be curious to infer stored data by analyzing the received queries and the stored data. Hospitals may be curious about the client's queries. Similarly, client may be curious about hospitals' sensitive information (genomic information and ASI stored at the CSP).  
In the proposed scheme, 
if the CSP and one of the  participating hospitals collude, the CSP may launch a brute-force attack to infer the SNP IDs and values contained in the indices of other hospitals. In this paper, we do not consider such a collusion. 
We briefly discuss the main threats against the proposed protocol in the following.



\noindent\textbf{Ciphertext attack}. The CSP  may attempt to infer the sensitive information of the hospitals by analyzing stored encrypted data.

\noindent\textbf{Query attack}. The CSP observes and processes the query from the client, and hence it may try to infer the query content (i.e., genomic data of the patient being queried). 

\noindent\textbf{Illegitimate access attack}. The client may try to access genomic data or ASI from a hospital's dataset without the authorization of the corresponding hospital.\vspace{-5pt}

\subsection{Query Model}
The query model is designed to provide the following functionality: given a (partial) sequence of SNPs representing a set of mutations for a patient, retrieve ASIs of patients whose mutations are similar to those of the given sequence. The input sequence does not need to include all mutations for a patient because the focus on the query can be on a specific pattern that includes a number of SNPs. The search is performed across the data from multiple hospitals under the constraint of access control. 

To prevent the CSP from learning the SNPs in the query input, the client transforms the input in the following way.
First, the client creates a Bloom filter and populates this filter with each input SNP. Then, the client extracts the positions of non-zero elements inside the Bloom filter and applies a pseudorandom function to each extracted position using a secret key as the seed. The result of this operation is used for the index search. 
Second, the client creates a query token for each input SNP by applying a hash function and encrypting the hash output with her private key. The outcome of this is used for the ASI search.
We present the details of these operations in Section~\ref{sec:cau} and discuss their privacy in Section~\ref{sec:secana}. 
We also let the client customize the search query by introducing two search parameters as follows:
\\
\textbf{The threshold for similarity metric ($\varepsilon_c)$}. Since each client $c$ may require a different level of similarity for a match, $c$ should be allowed to set its own minimum acceptable similarity value. In the proposed scheme, the threshold for similarity metric is equivalent to the minimum similarity score of the cosine metric. In the search phase, it is used to evaluate whether the match for a patient exceeds the threshold. 
\\
\textbf{The threshold for the number of retrieved results ($k_c$)}. The number of retrieved result is not predictable. Considering the constraints on the client's capacity, bandwidth, and personal preference, a client should be allowed to set the maximum number for the retrieved result. In the proposed scheme, this parameter is used to control the size of search result.

 \vspace{-5pt}
\section{Proposed Scheme}\label{sec:proposed}
\vspace{-5pt}


\subsection{Overview}

In order to perform similarity search, we use an index produced by hierarchical clustering. Building an index is an expensive procedure, which is performed by the cloud infrequently in an offline fashion. Once built, the index allows us to efficiently handle many queries on a daily basis. Due to privacy concerns, the cloud cannot build the index from scratch using the plaintext records. Therefore, a hospital first creates an encrypted index and sends it to the cloud, which subsequently performs hierarchical clustering. Additionally, the cloud needs to combine indices sent by different hospitals because a single query can search across multiple hospitals. The cloud can either keep and search individual hospital indices separately or merge them into a single index. Using individual indices may be more efficient when there are only a few hospitals whereas at a larger scale, it is better to merge indices.
The cloud uses a heuristic to determine which scheme to employ.

\begin{figure}[t]
    \centering
    \includegraphics[width=0.5\textwidth]{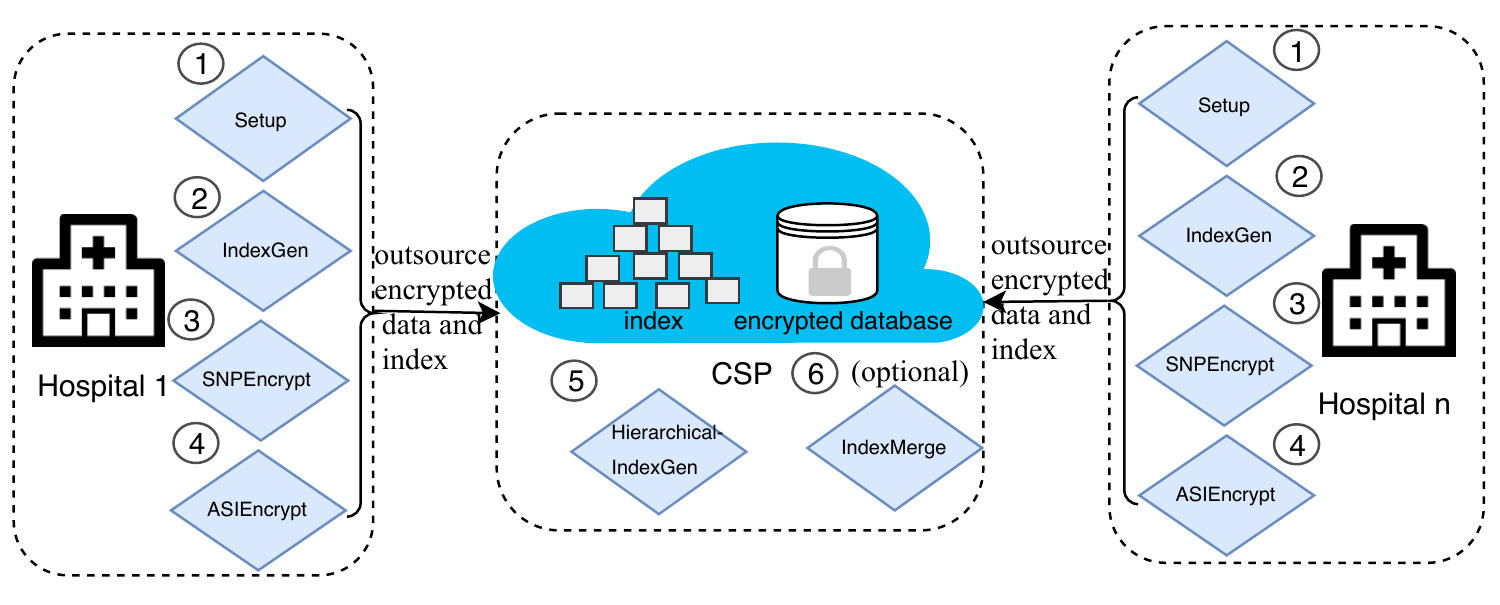}
    \caption{Initialization: The initialization process is conducted per hospital. It includes six modules: \emph{Setup}, \emph{IndexGen}, \emph{SNPEncrypt}, \emph{ASIEncrypt}, \emph{HierarchialIndexGen}, and \emph{IndexMerge}. The first four modules are sequentially executed by the hospital and output a privacy-preserving index and encrypted data. Each hospital sends its output to the CSP and the CSP calls \emph{HierarchicalIndexGen} to construct a hierarchical index based on the input. CSP calls  \emph{IndexMerge}  when the number of hierarchical indices exceeds a threshold.}
    \label{fig:initialisation}
    \vspace{-10pt}
\end{figure}
\begin{figure}[t]
    \centering
    \includegraphics[width=0.49\textwidth]{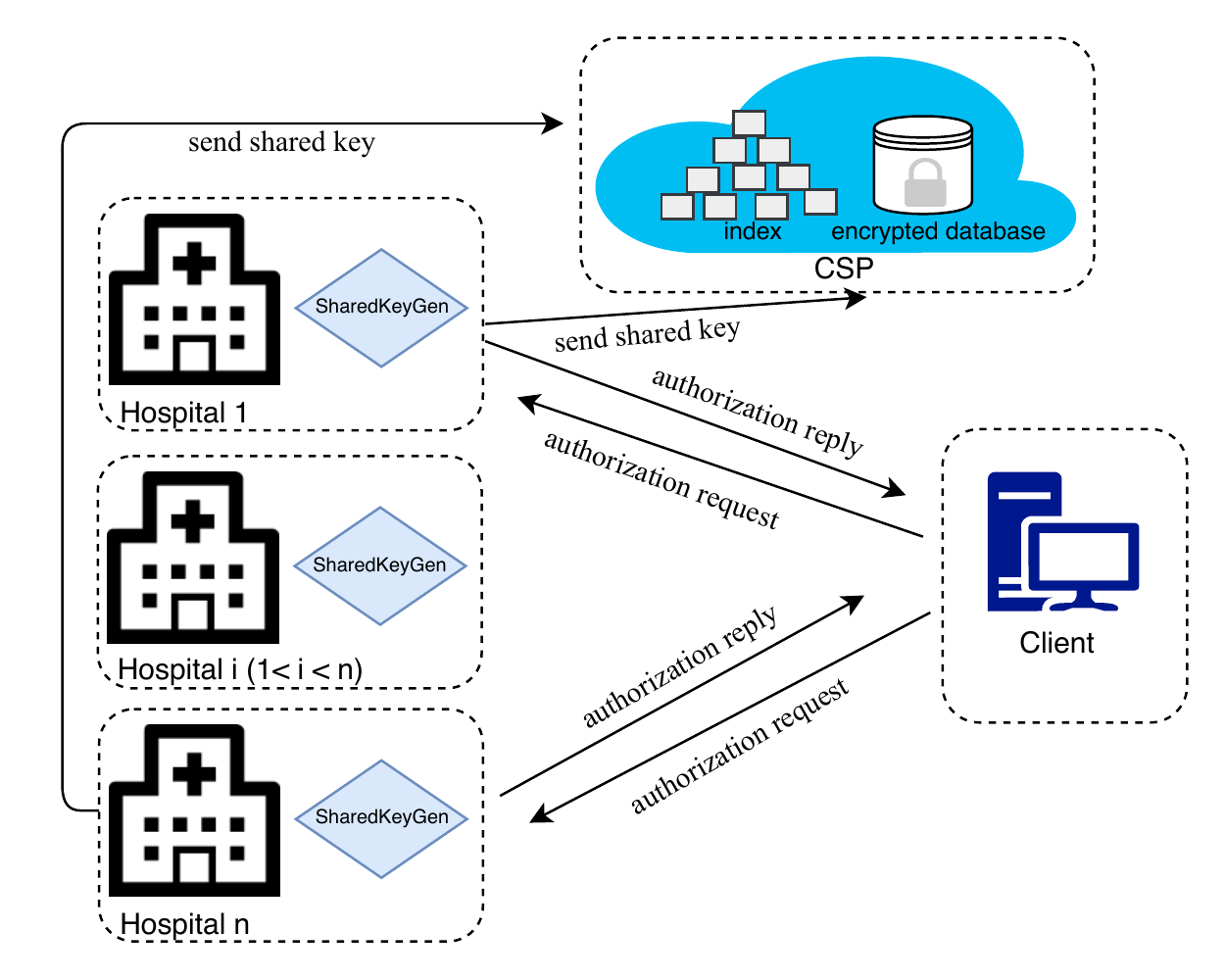}
    \caption{Client authorization: If a client wants to access a hospital's data, the authorization from the hospital is required. The client first sends an authorization request to the hospital. If the hospital approves the request, a token adjustment key is generated and sent to the CSP in addition to a successful authorization reply that is sent to the client.}
    \label{fig:clientauthorization}
    \vspace{-15pt}
\end{figure}
\begin{figure}[ht]
    \centering
    \includegraphics[width=0.48\textwidth]{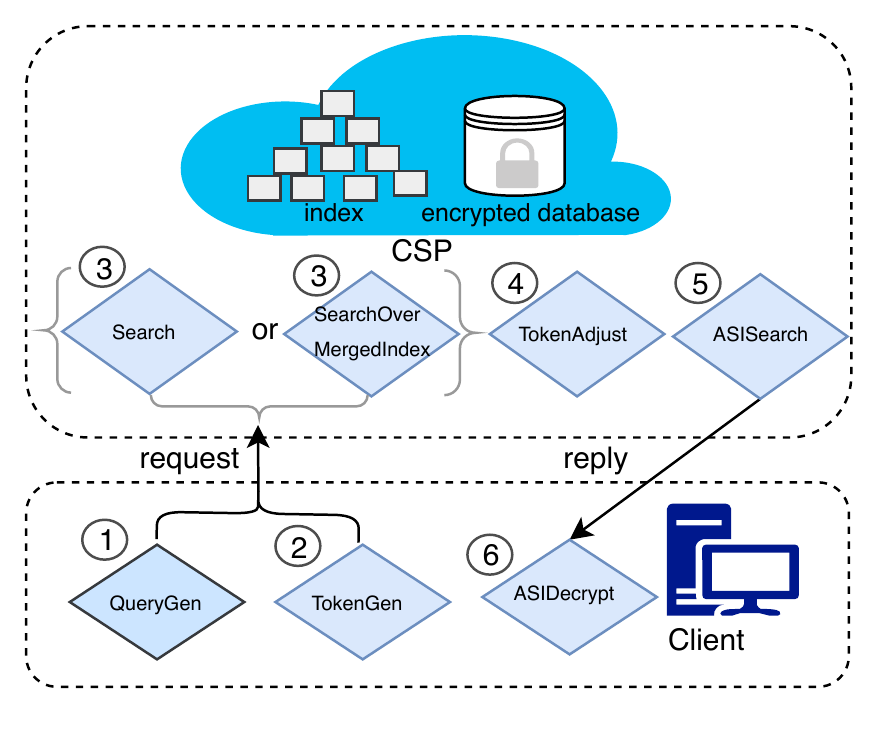}
    \caption{Query processing: A legitimate client first runs \emph{QueryGen} and \emph{TokenGen} to generate a query and sends it to the CSP. Upon receiving the query, the CSP runs either \emph{Search} or \emph{SearchOverMergedIndex} based on whether the CSP has merged the hierarchical indices. If the output is empty, the search process is terminated. Otherwise, the CSP calls \emph{TokenAdjust} to adjust tokens before \emph{ASISearch} is conducted. The outcome is sent to the client. The client runs \emph{ASIDecrypt} to obtain the plaintext ASIs of the similar patients.}
    \label{fig:queryprocessing}
    \vspace{-20pt}
\end{figure}

We divide the solution into three phases: initialization, client authorization, and query processing, as shown in Figures~\ref{fig:initialisation}, \ref{fig:clientauthorization}, and \ref{fig:queryprocessing}, respectively. The initialization is performed infrequently, depending how dynamic the system is.  The client authorization can be performed periodically, in line with common practices. The query processing is performed each time a user wants to send a search query. Each phase consists of a number of procedures, which are also summarized in Table~\ref{tab:csiss} in Appendix~\ref{sec:scsap}. We first present an overview of the procedures and then, provide their detailed descriptions.


In the initialization phase (shown in Figure~\ref{fig:initialisation}), each hospital first calls the \emph{Setup} function. \emph{Setup} chooses the initial parameters, configures library functions, and then preprocesses the dataset (e.g., by adding ASI to associated SNPs). 

After performing \emph{Setup}, each hospital runs the \emph{IndexGen} algorithm to build an index over its genomic data. The index is generated based on the genome similarity of its patients. Then, the encryption algorithms \emph{SNPEncrypt} and \emph{ASIEncrypt} are called to encrypt the genome sequences and corresponding ASIs.
To outsource the computation-intensive tasks to the CSP, the hospital directly sends the generated index and encrypted data to the CSP without building a hierarchial index over it. 
Upon receiving the encrypted index, the CSP runs the \emph{HierarchicalIndexGen} algorithm to build a hierarchical index in order to improve the search efficiency. Since each hospital outsources its indices to the CSP, the number of hierarchical indices stored at the CSP increases with the number of hospitals. If the number of hierarchical indices is beyond a certain threshold (that is determined by considering the efficiency of the search operation), the CSP calls the \emph{IndexMerge} function to merge all the hierarchical indices into one. 
We analyze the value of this threshold in Section~\ref{sec:perf}. 

If a client wants to perform a similar patient search, she first need to get authorization from a hospital before she can generate a valid query. Figure~\ref{fig:clientauthorization} shows the process of client authorization. The client sends an authorization request to a hospital. If the hospital approves the request, it generates a key for token adjustment and sends it to the CSP. Then, the hospital sends additional secret keys to the client, as described in Section~\ref{sec:cau}.

Once the client gets the authorization from a hospital, she can query the dataset of that hospital. An important advantage of our scheme is that the client can get an authorization from multiple hospitals and later send a single query to perform a search across all of them. Figure~\ref{fig:queryprocessing} shows the flow of query processing.
The client first calls the \emph{QueryGen} function to generate the first part of her query, which is used to search for the pseudonyms of the target similar patients. Then, the \emph{TokenGen} function is called to construct tokens as the second part of the query, which is used to retrieve the target ASIs. 
Token is constructed by encrypting client's input SNPs and it is adjusted using the token adjustment key (that is generated by the hospital and sent to the CSP). 
The adjusted token is used to provide controlled (or authorized) access to the client to the hospital's data. 
Upon receiving the query, the CSP first calls either \emph{Search} or \emph{SearchOverMergedIndex} function, depending on whether the algorithm \emph{IndexMerge} has been called or not, with the first part of the query to retrieve the pseudonyms of target similar patients. 

If the output of the first step is non-empty, the CSP adjusts the token using the second part of the query by running \emph{TokenAdjust}. Finally, the CSP calls the \emph{ASISearch} function to retrieve the ASIs belonging to the retrieved target patients. The result is sent back to the client. The client decrypts the received ciphertext by running \emph{ASIDecrypt} and obtains the plaintext ASIs of the corresponding target patients. 

\subsection{Initialization}
\label{sec:indini}
As shown in Figure~\ref{fig:initialisation}, the initialization consists of six modules: \emph{Setup}, \emph{IndexGen}, \emph{SNPEncrypt}, \emph{ASIEncrypt}, \emph{HierarchicalIndexGen}, and \emph{IndexMerge}. 
The first four modules are done at each hospital and the remaining modules are done at the CSP. 

To initialize the system, a trusted party that is in charge of key generation and distribution (such as the NIH) sets an asymmetric bilinear group ($G_1, G_2, G_T, p, e$), where $G_1$ and $G_2$ are two distinct groups of order $p$, and $e$ is the mapping from these two groups to the target group $G_T$. In addition, the settings (i.e., size and maximum false positive rate) of the standard Bloom filter \textit{BF} and customized Bloom filter \textit{CBF} are configured. After that, three hash functions ($H_0$, $H_1$, $H_2$) are chosen. 
These hash functions are used in the algorithms that will be discussed later.
Subsequently, a pseudorandom function (PRF) $F$ is chosen and a secret key $K$ is selected. We list the frequently used notations in Table~\ref{tab:ipf} (in Appendix~\ref{sec:keyp_func}).

\subsubsection{Setup (at the hospital) - \emph{Setup}}
In the first step of the \emph{Setup}, all the initial parameters and functions are selected. In the second step, dataset is preprocessed. 





The following steps are the same for all hospitals, and hence we describe them only for an hospital $i$. 
Hospital $i$ first generates two symmetric encryption keys, $K_{\alpha_i}$ and $K_{\beta_i}$ for the SNP encryption and ASI encryption, respectively. Then, it chooses a secret key $K_i$ for the shared key generation. Furthermore, a public/private key pair ($\textit{PK}_{i,1}, \textit{SK}_{i}$) is generated for the signature and a public/master key pair ($\textit{PK}_{i,2}, \textit{MK}_i$) is generated for CPABE.

\begin{figure}[t]
\centering 
\includegraphics[width=0.5\textwidth]{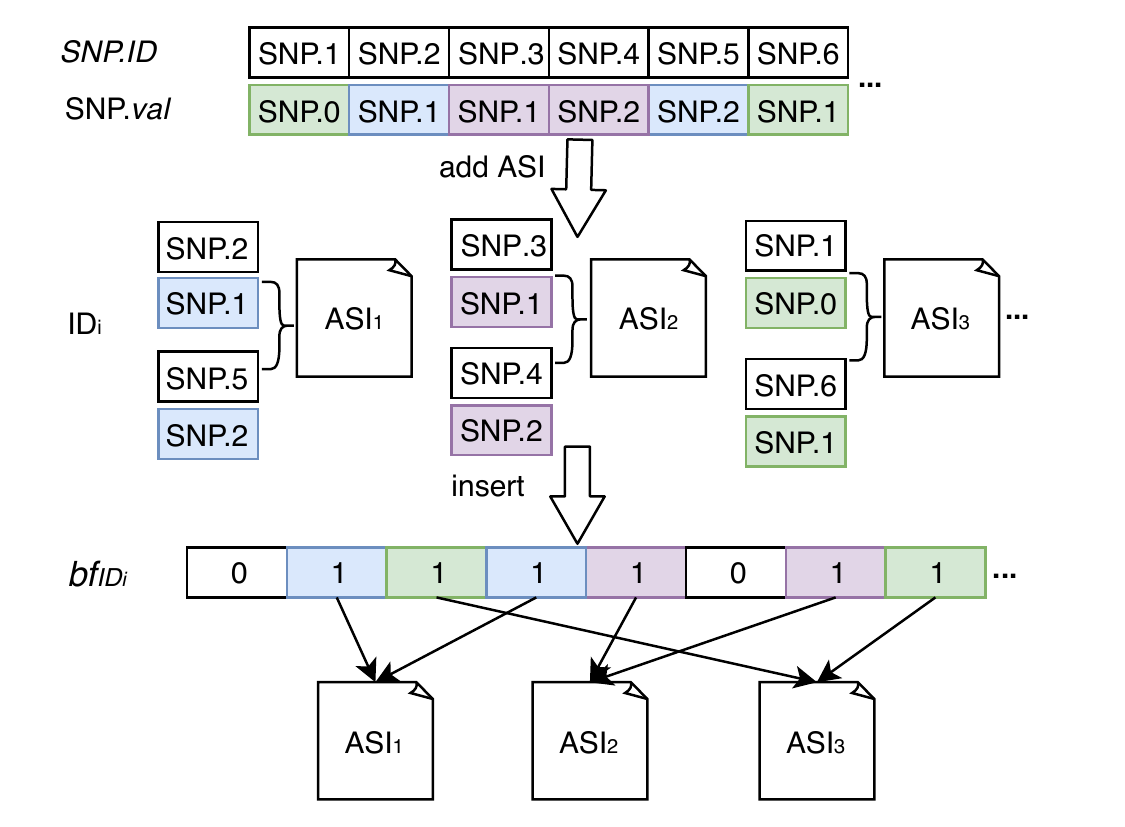}
\caption{Dataset preprocessing. The pair of SNP ID and value is first classified based on a certain property (e.g., its association with a disease) and then associated with an ASI. 
Afterwards, the concatenation of \textit{SNP.ID}  and \textit{SNP.val} is inserted into a Bloom filter. This generates the non-zero elements of the Bloom filter associated with ASIs.}
\label{fig:msb}
\vspace{-15pt}
\end{figure}

Next, the hospital starts preprocessing the dataset. 
The two phases of dataset preprocessing are shown in Figure~\ref{fig:msb}. 
In the first phase, the hospital adds a pseudonym for each patient record and ASIs for various groups of SNPs. The association between the ASIs and the SNPs can be determined based on several factors such as disease, phenotype, or treatment.
Eventually, data record belonging to a patient $\textit{ID}_i$ is represented as $\{ \textit{ID}_i$, $\{\mathbf{s}_{\textit{ID}_i,\textit{SNP}}^{\textit{ASI}_1}, \textit{ASI}_1)\}$, $\ldots$,  $\{\mathbf{s}_{\textit{ID}_i,\textit{SNP}}^{\textit{ASI}_k}, \textit{ASI}_k)\} \}$, where $\mathbf{s}_{\textit{ID}_i,\textit{SNP}}^{\textit{ASI}_k}$ represents a set of SNPs of individual $\textit{ID}_i$ (and their values) that are associated with $\textit{ASI}_k$. 
In the second phase, the concatenation of each SNP ID (represented by \textit{SNP.ID}) and SNP value (represented by \textit{SNP.val}) is mapped into a Bloom filter, denoted as $\textit{bf}_{\textit{ID}_i}$ (shown in Figure~\ref{fig:msb}). Eventually, this process generates the non-zero elements of the Bloom filter output that are associated with the ASIs of the corresponding patient. 

\renewcommand{\algorithmicrequire}{\textbf{Input:}}
\renewcommand{\algorithmicensure}{\textbf{Output:}}

\vspace{-5pt}
\subsubsection{Index generation (at the hospital) - \emph{IndexGen}}
\label{sec:indgen}

Each hospital indexes the records of its patients using a Bloom filter. The index generation algorithm is used to encrypt and randomize the non-zero elements of the Bloom filter. The same algorithm is also used to generate the query (as discussed in Section~\ref{sec:indexqg}).  
The details of the index generation (\emph{IndexGen}) algorithm are given in Appendix~\ref{app:initialization}. The input of the \emph{IndexGen} algorithm at hospital $i$ are the secret key $K$, the pseudorandom function (PRF) $F$, the dictionary $\textit{Dict}_i^\textit{BF}$, and the public/private key pair $(\textit{PK}_{i,1},\textit{SK}_{i})$. 
 
For each pseudonym $\textit{ID}_i$ in the dictionary $\textit{Dict}_i^\textit{BF}$, hospital $i$ connects the entry $\textit{Dict}_i^\textit{BF}[\textit{ID}]$ to the corresponding Bloom filter ($\textit{bf}_{\textit{ID}_i}$) that is constructed using the genome of patient with pseudonym $\textit{ID}_i$. 
If the value of a position $pos$ in the Bloom filter $\textit{bf}_{\textit{ID}_i}$ ($\textit{bf}_{\textit{ID}_i}[pos]$) is non-zero, then that position is extracted and input into the PRF $F$ with the secret key $K$. 
The CSP may understand if a patient exists in different hospitals' datasets and obtain more information about the patient in that way. 
To avoid this, the hospital selects a random string $r_i$ and invokes $F$ with the inputs $r_i$ and previous outcome of $F$. 
Since $r_i$ is a random string, the newly generated result 
is indistinguishable from a random input. 
The output is added into a customized Bloom filter $\textit{cbf}_{\textit{ID}_i}$. 
Once all the non-zero elements of the Bloom filter $\textit{bf}_{\textit{ID}_i}$ are mapped into the customized Bloom filter $\textit{cbf}_{\textit{ID}_i}$, 
the pair ($\textit{ID}_i, \textit{cbf}_{\textit{ID}_i}$) is added into the dictionary $\text{Dict}_i^\textit{CBF}$. 

\begin{figure}[t]
\centering
\includegraphics[width=0.48\textwidth]{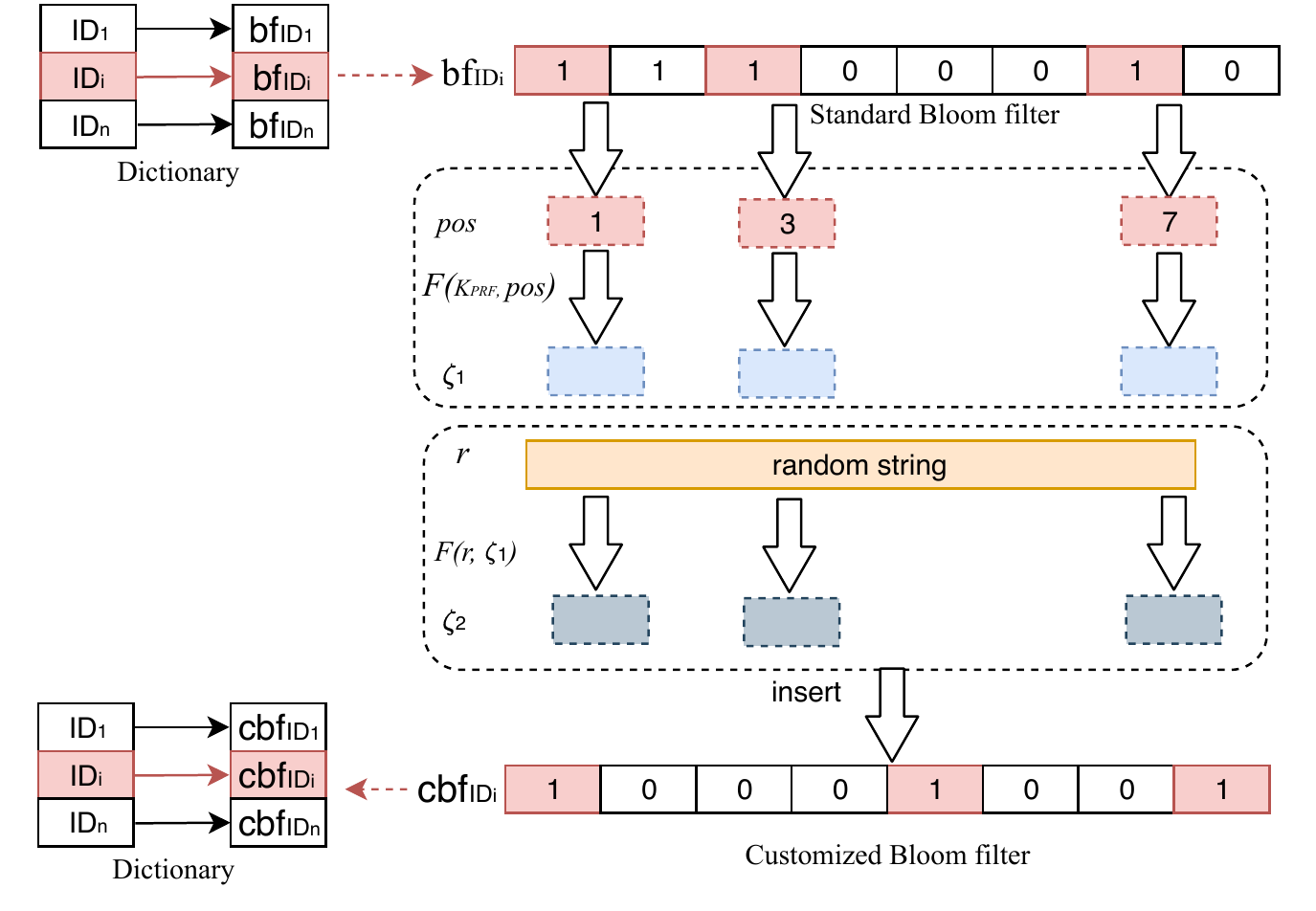}
\caption{Mapping from a standard Bloom filter to a customized Bloom filter. Given a standard Bloom filter, all the positions (\textit{pos}) of non-zero elements are extracted and encrypted by using a secret key as one of the inputs of a PRF.
To randomize the result ($\zeta_1$), a random  string ($r$) is selected and the PRF ($F$) is invoked anew with input of the random string and previous outcome. The result ($\zeta_2$)  is mapped into a customized Bloom filter.}
\label{fig:isa}
\vspace{-15pt}
\end{figure}

This process is also shown in Figure~\ref{fig:isa}. 
To verify the authenticity of the index, hospital $i$ digitally signs the hash of the concatenation of $\textit{PK}_{i,1}$ and $r_i$ by using $\textit{SK}_{i}$. 
The output of the algorithm is a tuple ($\Delta_i$, $r_i$, $\sigma_i$), where $\sigma_i$ is the signature, $r_i$ is the random string applied to build the index, and $\Delta_i$ consists of $\textit{Dict}_i^\textit{BF}$ and $\textit{PK}_{i,1}$.   

\vspace{-5pt}
\subsubsection{Data Encryption (at the hospital) - \emph{SNPEncrypt} and \emph{ASIEncrypt}}\label{sec:data_enc}
Data encryption consists of two parts. The first part is the encryption of the genome (i.e., SNPs) and the second part is the ASI encryption. 
For genome encryption, we propose \emph{SNPEncrypt} which utilizes the AES encryption algorithm. The input of the algorithm is the secret key $K_{\alpha_i}$ and a set $\mathbf{S}_{i,\textit{SNP}}$ of SNPs stored at hospital $i$. The output is a set $C_{i, \textit{SNP}}$ of encrypted genomes.

 
The ASI encryption algorithm at hospital $i$ includes two rounds of AES encryption with two different secret keys (as also shown in Figure~\ref{fig:asienc} in Appendix~\ref{sec:asienca}). 
In the first round, the secret key $K_{\beta_i}$ is used and in the second round, a secret key $K_{\gamma_i}$ that is randomly selected from the group $G_T$ is applied to encrypt the ciphertext from the first round. 
$K_{\beta_i}$ is held by hospital $i$ and it is only shared with the approved clients. $K_{\gamma_i}$ is encrypted using CPABE, which enables the access policy for the ASI. 
We describe the \emph{ASIEncrypt} algorithm in the following. The details of the algorithm are also given in Appendix~\ref{app:initialization}. 

The input of the algorithm includes two keys $K_i$ and $ K_{\beta_i}$ and a dictionary $\textit{Dict}_i^\textit{ASI}$.
Each item in the dictionary consists of two components. 
The first component is the pseudonym of a patient ($\textit{ID}_i$) and the second component is a list of ASIs belonging to the  patient. 
For each pseudonym $\textit{ID}_i$ in the dictionary $\textit{Dict}_i^\textit{ASI}$, hospital $i$ conducts the following operations. 
For each pair of ASI and $\mathbf{S}_{\textit{ID}_i,\textit{SNP}}^\textit{ASI}$ inside the $\textit{Dict}_i^\textit{ASI}[\textit{ID}_i]$, the hospital executes following four steps. 
First, for each SNP in set $\mathbf{S}_{ID_i, \textit{SNP}}^\textit{ASI}$, 
the hospital calls the hash function $H_2$ with a randomly selected value $\tau$ and bilinear mapping $e(H_1(v), g_2)^{1/K_i}$, where $v$ is the concatenation of \textit{SNP.ID} and its corresponding value \textit{SNP.val}.
The random value $\tau$ enables the hash result $h$ be indistinguishable from a random string. 
The result $h$ is added into a set $\theta$. 
Second, the AES encryption algorithm (\textit{AES.Enc}) is called to encrypt ASI with input key $K_{\beta_i}$ and it outputs the ciphertext $C_1$.
Third, a key $K_{\gamma_i}$ is chosen from $G_T$ and \textit{AES.Enc} is called again to encrypt $C_1$ with $K_{\gamma_i}$, resulting in ciphertext $C_2$.
Fourth, the secret key $K_{\gamma_i}$ is encrypted using CPABE with policy $\mathcal{A}$ built from the set $\theta$.
Specifically, all the elements inside $\theta$ are considered as attributes of the access policy.
The output ciphertext $C_3$ accompanied with $C_2$ and $\tau$ are added into a ciphertext set $\mathbf{C}_{ID_i}$.
After all the \textit{ASIs} of $\textit{Dict}_\textit{ASI}[\textit{ID}_i]$ are encrypted, the pair($\textit{ID}_i, \mathbf{C_{\textit{ID}_i}}$) is inserted into a dictionary $\textit{Dict}_i^C$.
Once this operation is done for all the patients (i.e., all the IDs have been processed), the algorithm outputs the dictionary $\textit{Dict}_i^C$. 

\vspace{-10pt}
\subsubsection{Hierarchical index generation (at the CSP) - \emph{HierarchicalIndexGen}}
\label{sec:hierarchicalindex}

\begin{figure}[t]
\centering
\includegraphics[width=0.5\textwidth]{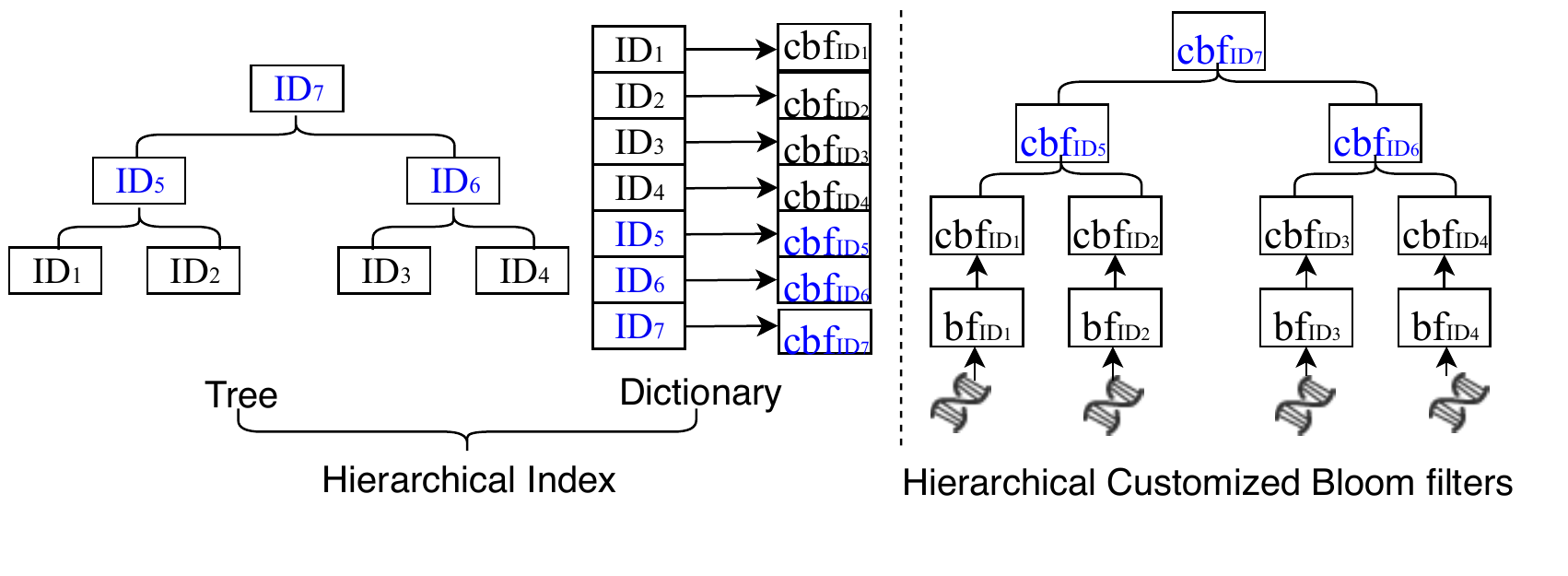}
\caption{Example of index structure and index construction process of hierarchical customized Bloom filters. $\textit{ID}_1$, $\textit{ID}_2$, $\textit{ID}_3$, $\textit{ID}_4$, $\textit{ID}_5$, $\textit{ID}_6$, and $\textit{ID}_7$ are pseudonyms, labelling a customized Bloom filter $\textit{cbf}_{\textit{ID}_i}$. 
The customized Bloom filter $\textit{cbf}_{\textit{ID}_i}$ is either directly or indirectly (colored blue) generated from a Bloom filter (or multiple Bloom filters). 
For example, $\textit{cbf}_{\textit{ID}_1}$ is mapped directly from $\textit{bf}_{\textit{ID}_1}$ while $\textit{cbf}_{\textit{ID}_5}$ is indirectly mapped from $\textit{bf}_{\textit{ID}_1}$ and $\textit{bf}_{\textit{ID}_2}$.}
\label{fig:eids}
\vspace{-10pt}
\end{figure}

The hierarchical clustering algorithm is designed to cluster the Bloom filters representing the genome sequences into hierarchical clusters. 
Also, a hierarchical index structure is designed to index all the hierarchical clusters with small memory requirement. 
The CSP builds hierarchical index based on the received (unclustered) index from each hospital $i$. 
The hierarchical index allows to search the target patient efficiently.
Figure~\ref{fig:eids} illustrates the hierarchical index structure and construction of hierarchical customized Bloom filters. The details of the algorithm are also given in Appendix~\ref{app:initialization}. 

The key part of the hierarchical clustering algorithm is setting the similarity metric, as it determines the quality of clustering. 
Instead of using traditional Euclidean distance as the similarity metric, in which one of the dimensions may be relatively large and may overpower the other dimensions, we choose the cosine similarity. Thus, in our protocol, the similarity metric is calculated as 
\textit{Sim}($\textit{cbf}_{\textit{ID}_i}$, $\textit{cbf}_{\textit{ID}_j}$)=$\frac{\textit{cbf}_{\textit{ID}_i}\cdot \textit{cbf}_{\textit{ID}_j}}{|\textit{cbf}_{\textit{ID}_i}|\cdot |\textit{cbf}_{\textit{ID}_j}|}$,
where $\textit{cbf}_{\textit{ID}_i}$ and $\textit{cbf}_{\textit{ID}_j}$ are two customized Bloom filters for patients $\textit{ID}_i$ and $\textit{ID}_j$, respectively. Also, $|\textit{cbf}_{\textit{ID}_i}|$ and $|\textit{cbf}_{\textit{ID}_j}|$ represent the lengths of $\textit{cbf}_{\textit{ID}_i}$ and $\textit{cbf}_{\textit{ID}_j}$. 
The inner product of Bloom filters $\textit{cbf}_{\textit{ID}_i}$ and $\textit{cbf}_{\textit{ID}_j}$ is equal to the sum of bitwise \emph{AND} of $\textit{cbf}_{\textit{ID}_i}$ and $\textit{cbf}_{\textit{ID}_j}$ since each element of a Bloom filter is either $0$ or $1$. The efficiency of computing the similarity score is enhanced by the bitwise operation. 

We use a similarity matrix to keep the pairwise similarity values between different Bloom filters representing genome sequences. 
Given the similarity matrix, the pairwise distances (i.e., 1-similarity value) can be easily calculated. Then, the multidimensional scaling (MDS) algorithm~\cite{borg2003modern} is invoked to compute the relative positions of genome sequences. Based on the relative positions, the classic hierarchical clustering algorithm is applied. For this, we use the classic Ward variance minimization algorithm~\cite{ward1963hierarchical}. 


We describe the \emph{HierarchicalIndexGen} algorithm below.  
The input of the algorithm is $\Delta_i$ consisting of $Dict_i^\textit{CBF}$, $\textit{PK}_{i,1}$, and $r_i$.
The CSP first extracts the dictionary $Dict_i^\textit{CBF}$ from $\Delta_i$. 
Then, the hierarchical clustering algorithm (denoted by \textit{HC}) is called with the input $Dict_i^\textit{CBF}$.
The outcome consists of a tree structure $\textit{Tr}_i$ and a new dictionary $Dict_i^{H}$.
Finally, the tree structure $\textit{Tr}_i$, dictionary $Dict_i^{H}$, public key $\textit{PK}_{i,1}$, and $r_i$ are gathered into a new tuple $\Delta_i^{H}$. 

In Figure~\ref{fig:eids}, we provide a toy example illustrating index structure and index construction process of hierarchical customized Bloom filters. We show $4$ patients, $7$ customized Bloom filters, and  $4$ standard Bloom filters. 
The customized Bloom filters, $\textit{cbf}_{\textit{ID}_1}$, $\textit{cbf}_{\textit{ID}_2}$, $\textit{cbf}_{\textit{ID}_3}$, $\textit{cbf}_{\textit{ID}_4}$, are constructed based on the standard Bloom filters (indirectly from patients' genomic data) by the hospital and sent to the CSP.
The remaining customized Bloom filters, $\textit{cbf}_{\textit{ID}_5}, \textit{cbf}_{\textit{ID}_6}, \textit{cbf}_{\textit{ID}_7}$, 
are generated by the CSP according to the tree structure, where $\textit{cbf}_{\textit{ID}_5}=
\textit{cbf}_{\textit{ID}_1} | \textit{cbf}_{\textit{ID}_2}$, $\textit{cbf}_{\textit{ID}_6}=
\textit{cbf}_{\textit{ID}_3} | \textit{cbf}_{\textit{ID}_4}$, and $\textit{cbf}_{\textit{ID}_7}= \textit{cbf}_{\textit{ID}_5} | \textit{cbf}_{\textit{ID}_6}$ ($|$ represents bitwise \textit{OR} operation). 
  


\vspace{-5pt}
\subsubsection{Index merging (at the CSP) - \emph{IndexMerge}} \label{sec:indim}

The index merging algorithm is invoked by the CSP once a high number of indices are received from different hospitals. The goal is to reconstruct an efficient index to replace all the stored indices without any loss in terms of utility and privacy. 
 
We describe the \emph{IndexMerge} algorithm in the following. The details of the algorithm are also given in Appendix~\ref{app:initialization}.
The CSP first initializes a temporary dictionary $Dict_\textit{temp}$ 
and a set $\mathbf{\Upsilon}$. $Dict_\textit{temp}$ is applied to store the pair of pseudonym and corresponding customized Bloom filter. 
The set $\mathbf{\Upsilon}$ is used to store pairs including a public key and the corresponding random string. 
For each index, the CSP resolves it into a tree $\textit{Tr}_i$, a dictionary $Dict_i^\textit{CBF}$, a public key $\textit{PK}_{i,1}$, and a string $r_i$. $\textit{PK}_{i,1}$ and $r_i$ are collected into the set $\Upsilon$. Then, each leaf node \textit{n} of $\textit{Tr}_i$ is read and the value of entry $Dict_i^\textit{CBF}[\textit{n.ID}_i]$ is added into the $Dict_\textit{temp}$ with the entry constructed by the concatenation of $\textit{n.ID}_i$ and $r_i$. 
The reason of concatenating $\textit{n.ID}_i$ and $r_i$ is to avoid the same pseudonym appearing in different hospitals.  
After all the indices are processed, the CSP runs hierarchical clustering algorithm \textit{HC} over the $Dict_\textit{temp}$ and outputs the new tree structure $\textit{Tr}_{M}$ and dictionary $Dict_{M}^\textit{BF}$. The output of the algorithm is a merged index \textit{MI} consisting of $\mathbf{\Upsilon}$, $\textit{Tr}_{M}$, and $Dict_{M}^\textit{CBF}$.

An alternative faster approach for \emph{IndexMerge}
is to  merge the roots of the hierarchical indices instead of merging all the leaves from scratch.
In this way, in the new hierarchical index, each leaf becomes a root of the original hierarchical index. 
The advantage of this fast approach is the reduced time to build the new index.
Using the faster approach, the time complexity of \emph{IndexMerge} decreases from $O(N\log N)$ to $O(n\log n)$, where $N$ is the total number of leaves and $n$ is the total number of roots.
The disadvantage is that the new index does not precisely cluster all the similar nodes into a cluster across hospitals. This may result in increased search time since the time complexity of search is determined by the search path from the root of the index to all the matching leaves.
We implement and evaluate the performance of this faster approach in Section~\ref{sec:perf}.
\vspace{-5pt}

\vspace{-5pt}
\subsection{Client Authorization}\label{sec:cau}

Upon a hospital $i$ receives an authorization request from a client, the hospital makes a decision on whether to allow its genomic data to be accessed or not. If the hospital approves the request, a shared key is generated and sent to the CSP and a success message is sent back to the client along with the keys $K$ and $K_{\beta_i}$.
Otherwise, a failure message is sent to the client. The shared key is used by the CSP to transform the client's query token into a searchable token over ASI. We describe the shared key generation algorithm in the following. The details of the algorithm are also given in Appendix~\ref{app:client_auth}.
The input of the algorithm includes two keys $K_i$ and $K_c$ and a set $\mathbf{S}_{i,\textit{SNP}}$ of SNPs. $K_i$ and $\mathbf{S}_{i,\textit{SNP}}$ are from hospital $i$, while $K_c$ is from client $c$. $\mathbf{S}_{i,\textit{SNP}}$ is the set of SNPs that the hospital allows the client to access. 
The hospital generates the first part of the shared key ($\delta_i$) by computing $g_2^{K_c/K_i}$. 
Then, for each SNP in $\mathbf{S}_{i,\textit{SNP}}$, the hospital executes the following two procedures. First, the hash function $H_1$ is invoked with the input SNP and the hash result is raised to the power $1/K_c$. Second, the previous outcome is added into a customized Bloom filter $\textit{cbf}_i$.
The final output is the shared key consisting of two parts: $\delta_i$ and $\textit{cbf}_i$.

\vspace{-5pt}
\subsection{Query Processing}\label{sec:query_processing}
As shown in Figure~\ref{fig:queryprocessing}, query processing includes: query generation (\emph{QueryGen} and \emph{TokenGen}), search (\emph{Search}, \emph{SerachOverMergedIndex}, \emph{TokenAdjust}, \emph{ASISearch}), and decryption (\emph{ASIDecrypt}).

\vspace{-5pt}
\subsubsection{Query generation - \emph{QueryGen} and \emph{TokenGen}}\label{sec:indexqg}

Query generation is executed by a client. It consists of two algorithms, \emph{QueryGen} and \emph{TokenGen}. The outcome of the \emph{QueryGen} is used to search over indices of target pseudonyms, while the outcome of the \emph{TokenGen} is used to search over the encrypted ASIs.


\emph{QueryGen} algorithm is detailed in Appendix~\ref{app:query_process}. 
The input of the algorithm for a client $c$ consists of the secret key $K$, a set $\mathbf{S}_{c,\textit{SNP}}$ of SNPs, a threshold $\varepsilon_c$ representing the minimum number of matching SNPs for a successful search, and a parameter $k_c$ to specify the threshold for the maximum number of retrieved pseudonyms. 
Each pair of \textit{SNP.ID} and \textit{SNP.val} inside $\mathbf{S}_{c,\textit{SNP}}$ is concatenated and added into a Bloom filter $\textit{bf}_c$. 
The procedure is same as the process in index generation (as in Section~\ref{sec:indgen}). 
For each non-zero element inside the $\textit{bf}_c$, the position $pos$ is extracted and computed by calling $F$ using the secret key $K$ as the input. The output is added into a set $\mathbf{E}$. 
The threshold $\varepsilon_c$ is set by computing the number of non-zero bits caused by the minimum number of SNPs. That is, $\frac{\textit{count}}{|S|}\cdot \varepsilon_c$, where \textit{count} is the total number of non-zero bits caused by the input SNPs and $|S|$ is the total number of input SNPs. 
The final form of the output is a tuple ($\mathbf{E}, \varepsilon_c, k_c, \sigma_i$), where signature $\sigma_i$ is generated by the hospital $i$ and sent to an approved client.



The details of the \emph{TokenGen} algorithm are also given in Appendix~\ref{app:query_process}. The input of the \emph{TokenGen} algorithm includes a secret key $K_c$ and a set $\mathbf{S}_{c,\textit{SNP}}$ of SNPs. For each SNP inside $\mathbf{S}_{c,\textit{SNP}}$, a hash function $H_1$ is called and the hash result is raised to the power $1/K_c$ for encryption and future token adjustment. 
Each outcome is collected into a set $\mathbf{TK}$. Finally, the algorithm outputs $\mathbf{TK}$. 

Eventually, the query sent to the CSP consists of the outputs of algorithms \emph{QueryGen} and \emph{TokenGen}. 

\vspace{-5pt}
\subsubsection{Search over a single index - \emph{Search}}\label{sec:indsis}

The search algorithm is run by the CSP. 
For clarity of the presentation, we first consider a scenario in which the search algorithm runs over a single index (belonging to a single hospital).
The search algorithm only uses the first part $Q$ of the query (encrypted input SNPs~$\mathbf{E}$) to traverse the index from the root to the leaves considering the minimum similarity threshold $\varepsilon_c$. The details are explained as follows (the details of \emph{Search} algorithm are also given in Appendix~\ref{app:query_process}).

The CSP receives a query $Q$ from a client $c$ and an index ($\Delta_i, r_i$) from a hospital $i$. 
The CSP first resolves $\Delta_i$ into a tree $\textit{Tr}_i$, a dictionary $Dict_i^{H}$ (which consists of pairs of patient pseudonym and corresponding customized Bloom filter), and a public key $\textit{PK}_{i,1}$. Query $Q$ is resolved into a set $\mathbf{E}$, a signature $\sigma_i$, and two threshold values $\varepsilon_c$ and $k_c$.
Then, the CSP builds a dictionary $Dict$ with only $k_c$ entries. 
The dictionary $Dict$ stores pairs including (i) patient pseudonym and (ii) similarity score between corresponding patient's customized Bloom filter and the queried genome's customized Bloom filter. 

Afterwards, the CSP verifies $\sigma_i$ by running the verification function \textit{verify} with the inputs $\textit{PK}_{i,1}$, $\sigma_i$, and $H(\textit{PK}_{i,1}, r_i)$.
If the verification fails, the process is terminated. 
Otherwise, the CSP continues to execute the following procedures. 
For each element $\zeta_1$ in the set $\mathbf{E}$, the CSP runs $F$ with the inputs $\zeta_1$ and $r_i$. The outcome $\zeta_2$ is added into a customized Bloom filter $\textit{cbf}_c$. 
After completing the above process, the CSP reads the root ($\textit{root}_i$) from $\textit{Tr}_i$ and pushes it into the queue \textit{qu}.   

Following steps are recursively executed until the \textit{qu} is empty. 
First, a node (\emph{n}) is popped out from the \textit{qu}. 
Second, the similarity score \textit{sim} is computed by using the cosine similarity between $Dict[\textit{n.ID}_i]$ and $\textit{cbf}_c$.
If $sim$ is less than $\varepsilon_c$, then the next step is skipped and step one is invoked again. 
If $sim$ is greater than or equal to $\varepsilon_c$, the property of \emph{n} is checked. 
If \emph{n} is a leaf, we call the insert function (\emph{Insert}) with the inputs $Dict$, $\textit{n.ID}_i$, and $sim$. 
The details of \emph{Insert} are given in Appendix~\ref{app:query_process}. The purpose of \textit{Insert} function is to insert the pair ($\textit{ID}_i$, \textit{sim}) into \textit{Dict} if \textit{Dict} is not full or there exits a pair that has smaller similarity score compared to the current node. 
If \emph{n} is not a leaf and there exists a left child (\textit{leftchild}), then \textit{leftchild} is pushed into $qu$. 
If its right child (\textit{rightchild}) exits, then \textit{rightchild} is pushed into $qu$. 
After this iteration is completed, the CSP outputs the final result \emph{Dict}.

\vspace{-5pt}
\subsubsection{Search over a merged index - \emph{SerachOverMergedIndex}}
\label{sec:indsomi}

Compared to searching over a single index, searching over a merged index mainly differs in two aspects.
First, the CSP verifies all the signatures submitted by the client instead of a single one (to recognize which hospitals in the merged index authorize the search). 
Second, each random string attached to an authorized index is used to generate a customized Bloom filter based on the submitted query. 

We describe the \emph{SearchOverMergedIndex} algorithm in the following. The details of the algorithm are also given in Appendix~\ref{app:query_process}. 
The input of the \emph{SerachOverMergedIndex} algorithm is a merged index $\varrho$ and a query $Q$ from a client $c$.
The query $Q$ includes a set $\mathbf{E}$ of encrypted SNPs, a set $\sigma_{s}$ of signatures, and two thresholds $\varepsilon_c$ and $k_c$. The merged index is resolved into a set $\mathbf{\Upsilon}$, a tree structure $\textit{Tr}_M$, and a dictionary $Dict_M^\textit{BF}$.
For each signature $\sigma_i \in \mathbf{\sigma}_{s}$, the CSP verifies whether there exists a pair $(\textit{PK}_{i,1}, r_i) \in \mathbf{\Upsilon}$ that matches $\textit{verify(PK}_{i,1}, \sigma_i, H(\textit{PK}_{i,1},r_i))=\textit{True}$.  
For each pair of $(\textit{PK}_{i,1}, r_i)$ that matches the verification, the random string $r_i$ is extracted and input into $F$ with each $\zeta_1$ in $\mathbf{E}$. 
The output $\zeta_2$ of $F$ is added into the customized Bloom filter $\textit{cbf}_c$.  
If the customized Bloom filter is empty, it means no submitted signature is valid and the algorithm returns \emph{None}. Otherwise, $\textit{cbf}_c$ is used to search over the tree \emph{$\textit{Tr}_M$}. The process is similar to algorithm \emph{Search} in Section~\ref{sec:indsis} (details are shown in Appendix~\ref{app:query_process}). The only difference is that the \emph{Insert} function is replaced by \emph{InsertConditionally}. 
The details of \emph{InsertConditionally} are also given in Appendix~\ref{app:query_process}.
Compared with \emph{Insert}, the difference is that in the \emph{InsertConditionally} algorithm, the input pseudonym $\textit{ID}_i$ is the concatenation of a real patient pseudonym and a random string that is required to be inside the authorized set.  
The extra operation is to verify the legitimacy of the record and to guarantee that all the records stored in $Dict$ are authorized. 
Finally, the \emph{SearchOverMergedIndex} algorithm outputs $Dict$.

\vspace{-5pt}
\subsubsection{Token adjustment - \emph{TokenAdjust}}
\label{sec:asita}

The token received from the client is not directly applicable for search over the ASI ciphertext. 
The CSP needs to use a shared key to transform the received token into an executable token (see Appendix~\ref{app:query_process} for details).
The input of the \emph{TokenAdjust} algorithm consists of a set of tokens ($\mathbf{TK}$) and a shared key $(\delta_i, \textit{cbf}_i)$ from a hospital $i$.
For each token $tk$ in $\mathbf{TK}$, the membership evaluation is conducted over the customized Bloom filter $\textit{cbf}_i$. 
If $\textit{cbf}_i(tk) \neq \textit{False}$, the bilinear mapping algorithm $e$ is called with inputs $tk$ and $\delta_i$. The result is collected into a set $\mathbf{TK}^*$. Otherwise, the current round of $tk$ is skipped. Finally, the algorithm outputs $\mathbf{TK}^*$, which can be used to search the target ASIs.

\vspace{-5pt}
\subsubsection{ASI search - \emph{ASISearch}} \label{sec:asis}

Given executable tokens (for search) and ciphertext, the CSP can proceed with ASI search. To identify an ASI, the tokens must match the access policy set on the ASI.
The access policy is set in the granularity of SNPs. 
For example, ASIs related to the diagnosis and treatment of breast cancer are encrypted by considering the SNPs of BRCA gene as the required attributes.   
That is, these attributes are applied to construct a polynomial that outputs the value of a target parameter (secret key) when all the required attributes are satisfied. 


The details of ASI search are also given in Appendix~\ref{app:query_process}. 
The input of the ASI search algorithm includes a set $\mathbf{TK}^*$ of executable tokens,
a set $\mathbf{\eth}$ of pseudonyms of patient records obtained from index search, a dictionary $Dict_i^\textit{C}$ containing pseudonym and ASI ciphertext pairs, and a master key $\textit{MK}_i$ of CPABE. 
For each pseudonym $\textit{ID}_i$ in $\mathbf{\eth}$, the entry
$Dict_\textit{ASI}[\textit{ID}_i]$ associcates ciphertexts $C_2$, $C_3$ and a random string $\tau$. 
To decrypt ciphertext $C_2$, the secret key ($sk$) of CPABE is generated as follows. 
For each token $tk^*$ in $\mathbf{TK}^*$, the hash function $H_2$ is called with the inputs $\tau$ and $tk^*$.
The result is gathered into a set $\mathbf{\vartheta}$. 
Then, the key generation algorithm \textit{CPABE.KeyGen} is called with the master key $\textit{MK}_i$ of CPABE and $\mathbf{\vartheta}$. 
If the attribute set $\mathbf{\vartheta}$ does not match the access policy, the newly generated $sk$ is null and following operations are skipped. 
Otherwise, the following procedures are executed to open the first layer of ciphertext $C_2$. 
The decryption algorithm (\textit{CPABE.Decrypt}) of CPABE is invoked with the inputs $sk$ and $C_3$.
If the output symmetric key $K_{\gamma_i}$ is not null, the decryption algorithm \textit{AES.Decrypt} of AES is invoked with inputs $K_{\gamma_i}$ and $C_2$.  
The output $C_1$ is collected into a set $\mathbf{C}_{C_1}$. 
Once all the elements inside $\mathbf{\eth}$ are accessed, 
the search algorithm outputs the set $\mathbf{C}_{C_1}$ that will be sent to the client.

\vspace{-5pt}
\subsubsection{ASI decryption - \emph{ASIDecrypt}}
\label{sec:asid}

Upon receiving the search result, the client applies its secret key to decrypt the ciphertext of ASI. 
For each ciphertext of retrieved result $\mathbf{C}_{C_1}$, the decryption algorithm of AES is invoked to decrypt the ciphertext with the input secret key $K_{\beta_i}$ assigned by hospital $i$. The plaintext \textit{ASI} is gathered into a set $\mathbf{S}_\textit{ASI}$. After all the ciphertext is decrypted, the algorithm outputs $\mathbf{S}_\textit{ASI}$. The details of this operation are also given in Appendix~\ref{app:query_process}.

\vspace{-5pt}
\section{Privacy Analysis}\label{sec:secana}
\vspace{-5pt}

In this section, based on the threat model described in Section~\ref{ssec:tm}, we prove that our scheme meets the privacy goals. Following previous work~\cite{chase2010structured, curtmola2011searchable,zhu2019privacy}, we consider the following as the allowed leaked information to the adversary throughout the protocol: (i) the size pattern, (ii) search pattern, and (iii) access pattern. Based on this information, we define a ``leakage function'' formalizing the information that is allowed to be learnt by the adversary. We provide the details of the leakage function in Appendix~\ref{sec:leak_func}.

\vspace{-5pt}
\subsection{Privacy Definition}
The privacy of the proposed scheme consists of two components. The first is the privacy of genomic data and the second is the privacy of ASI. 
Privacy of genomic data can be further split into index privacy and privacy of genome sequences. 
Genome sequences are encrypted using AES and they are not involved in the query processing.
Thus, their privacy relies on the robustness of AES. 
Since AES encryption achieves semantic security (e.g., CBC and CTR modes~\cite{lipmaa2000comments}), the encrypted genome sequences are robust against chosen plaintext attacks. 
The index privacy depends on the probability of reversing the customized Bloom filter to obtain the genome sequences using the information from the query execution and analyzing the customized Bloom filter. 
We formulate the privacy of genomic data as a game between a challenger and an adversary. First, the adversary selects two datasets $DB_{0,SNP}$ and $DB_{1, SNP}$ of SNPs. Each item of dataset $DB_{i, SNP}~(i \in \{0,1\})$ includes two components: patient pseudonym and corresponding SNPs of the patient. Then, the adversary sends the two databases to the challenger. The adversary is allowed to send adaptive queries with constraint on the information leakage before making the final decision about which dataset is utilized by the challenger. We provide the details of this game in  Appendix~\ref{sec:genomic_privacy}.
Similar to privacy of genomic data, we also define the privacy of ASI via a game between a challenger and the adversary. The formal definition of this game is in  Appendix~\ref{app:ASI_privacy_game}. 


\vspace{-5pt}
\subsection{Privacy Proof}
For the proof of genomic data privacy, we show that in the genomic data privacy game (in Appendix~\ref{sec:genomic_privacy}), $DB_{0, SNP}$ and $DB_{1,SNP}$ are indistinguishable for a PPT adversary.\vspace{-3pt}

\begin{theorem}
\label{thr:indone}
Let $\Pi_{SNP}$=\{\emph{Setup}, \emph{IndexGen}, \emph{QueryGen}, \emph{IndexMerge}, \emph{Search}, \emph{SearchOverMergedIndex}\} be a set of algorithms of the proposed scheme that are related to genomic data privacy. The scheme $\Pi_{SNP}$ is privacy-preserving if no PPT adversary can distinguish  $DB_{0,SNP}$ from $DB_{1,SNP}$ with non-negligible advantage at the end of the game defined in Appendix~\ref{sec:genomic_privacy}.\vspace{-3pt}
\end{theorem}

It is trivial to verify that the above theorem is consistent with the privacy definition of genomic data. This leads us to the following theorem.\vspace{-3pt}
 
\begin{theorem}
\label{thr:indtwo}
A PPT adversary cannot distinguish the view of $DB_{0,SNP}$ from the view of $DB_{1, SNP}$ in the game defined in Appendix~\ref{sec:genomic_privacy} if the applied AES encryption mode (e.g., CBC and CTR) is semantically-secure and PRF is indistinguishable from a random function. \vspace{-3pt}
\end{theorem} 

We provide the proof of Theorem~\ref{thr:indtwo} in Appendix~\ref{app:thorem_indtwo}. According to Theorems~\ref{thr:indone} and~\ref{thr:indtwo}, we conclude that the scheme $\Pi_{SNP}$ is privacy-preserving. The privacy of the ASI can be also proved similarly. 

\vspace{-5pt}
\section{Evaluation}\label{sec:perf}
\vspace{-5pt}

In this section, we evaluate the efficiency and scalability of the proposed scheme. 
Since the memory mapping technique~\cite{lea1996memory, dulloor2014system, lin2014mmap} is widely applied in our implementation of the proposed scheme (e.g., small part of the index is stored in memory and the remaining is stored on the disk), a stable memory address is required to measure the memory usage for different scenarios. Due to this requirement, we first conducted experiments on a single machine to show the efficiency of the proposed scheme and to analyze the run-time under different scenarios.
Due to the resource constraints of a single machine, we then turned to Amazon EC2 platform~\cite{awsec2} for running large-scale experiments to show the scalability of the proposed scheme.
In all experiments, the length of security parameter of RSA signature is set to $3096$ bits. We use RSA only for the digital signature and it is only computed once per query by the data owner. Thus, it has a negligible effect on the overall performance.
Index is built with a key ($K$) of size $256$ bits and the genomic data is encrypted using AES with a key ($K_{\beta_i}$) of size $256$ bits. 
The asymmetric curve used in bilinear mapping is set to \emph{MNT224} and the symmetric curve applied in CPABE is set to \emph{SS512}. The bilinear mapping is applied when the authorization protocol runs, while the CPABE is called for the ASI encryption. In our implementation, both MNT224 and SS512 provide $90$ bits of security. 
The capacity of Bloom filter is set to $2^{21}$, and maximum false positive rate of the Bloom filter is set to $0.01$. 
Also, in all experiments, we generated each query by setting the threshold for the minimum number of matching SNPs ($\varepsilon_c$) to 90 percent of the total input SNPs and the threshold for the number of retrieved results ($k_c$) to $5$. In addition, the number of SNPs in a query is equal to the number of SNPs per patient, if not specified otherwise.  
The client is assumed to be authorized to access all the hospitals' data. 
Finally, we run each experiment 10 times and report the average performance.

\vspace{-5pt}
\subsection{Experiments on a Single Local Machine}  
We ran the single machine experiments using a computer with Ubuntu system, i7 CPU, 32GB RAM, and 500GB hard disk.

\vspace{-5pt}
\subsubsection{Data Model}
We used the \emph{rsnps} tool~\cite{rsnps} to obtain all the raw patient files from the publicly available OpenSNP dataset~\cite{OpenSNP}. The whole dataset includes 3477 individuals and its plaintext size is 55GB. We first converted the raw patient files into VCF format using an open source software called \emph{personal-genome-analysis}~\cite{oga}. Eventually, we ended up with 2850 valid VCF files. 
For the affiliated sensitive information (ASI), we also used the OpenSNP dataset. In total, we collected 7388 ASIs and we randomly assigned them to the patients in varying numbers. The number of SNPs associated to an ASI varies from 20 to 2000.

\vspace{-5pt}
\subsubsection{Results}

In our Bloom filter settings, the false positive rate is 1 percent for an input size of 2 million SNPs. 
When the input size increases to 3 million SNPs, the false positive rate increases to 6 percent. 
However, in our experiments, we did not observe such a high accuracy loss. The reason is that in the dataset few patient records had 3 million SNPs. Most patient records had around 2 million SNPs, so that the precision is at least 99 percent in all experiments.


We first evaluated the proposed mechanism when a hospital has 10, 100, 1000, and 2850 patients and each patient has 20 SNPs. 
The results are shown in Table~\ref{tab:igdone}. We observed that the time cost of \emph{QueryGen} is constant, while the time costs of \emph{IndexGen} and \emph{Search} increases linearly with the increasing patient records.
Furthermore, the growth rate of the time cost of \emph{HierarchicalIndexGen} is approximately equal to the square of the growth rate of the patient records. 
In addition, the memory and disk storage requirements increase linearly with the number of patients. 

{
\footnotesize
\begin{table}
 \centering
  \caption{Performance (in terms of time cost in seconds and storage cost) of the proposed scheme with different number of patients in a hospital's database. In all scenarios, each patient has 20 SNPs.} 
 \label{tab:igdone}
 \begin{tabular}{|l|c|c|c|c|}
 \hline 
 Number of patient records & 10 & 100 & 1000 & 2850 \\
 \hline 
 IndexGen (s)& 0.001 &  0.01 & 0.065   & 0.188   \\
 \hline
 HierarchicalIndexGen (s): &0.24 & 23.5 & 2351 & 19772 \\
 \hline
 QueryGen (s) & 0.005 & 0.005 & 0.005 & 0.005 \\
 \hline 
 Search (s)&  0.037 &  0.386  & 3.858   & 10.036   \\
 \hline
Index Size in RAM (B) & 7.4K & 47.3K& 355K & 1005K   \\
  \hline 
Index Size in Disk (B) &  45.5M  & 476.8M   & 4.68G & 13.3G   \\
 \hline
Query Size  (B) & 8.28K & 8.28K& 8.28K &8.28K  \\
\hline 
 \end{tabular}
 \vspace{-10pt}
 \end{table}
}

Then, we considered the scenario that includes a hospital with 2850 patients and each patient having 200, 2000, and 3350221 SNPs (whole sequence), respectively. The results are shown in Table~\ref{tab:igdtwo}.
We observed that the time costs of \emph{IndexGen} and \emph{QueryGen} algorithms increase almost linearly with the increasing number of SNPs per patient. 
We also observed that the time costs of \emph{Search} and \emph{HierarchicalIndexGen} algorithms do not strongly correlate to the number of SNPs per patient; there is only slight increase in time cost when the number of SNPs increases dramatically. 
The index size is independent of number of SNPs. However, the size of the query increases linearly with the increasing number of used SNPs. 

 {
 \footnotesize
\begin{table}
 \caption{Performance (in terms of time cost in seconds and storage cost) of the proposed scheme with different number of SNPs per patient. In all scenarios, the hospital has 2850 patients.}
   \label{tab:igdtwo}
 \centering
 \begin{tabular}{|l|c|c|c|}
 \hline 
  Number of SNPs per patient  & 200 &2000 & 3350221\\
 \hline 
 IndexGen (s)  &  0.65 & 2.92&    175.86     \\
 \hline
 HierarchicalIndexGen (s): & 19775 & 19799  &   19936\\
 \hline
  QueryGen (s) & 0.005 & 0.052 &   2.92  \\
 \hline 
 Search (s)&   10.93 & 11.00 &  11.59   \\ 
  \hline
  Index Size in the RAM (B) &1005K& 1005K& 1005K \\
  \hline 
  Index Size in the Disk (B) & 13.3G & 13.3G & 13.3G \\
 \hline
 Query Size (B)  & 51.3K & 482K&  961883.35K \\
 \hline
 \end{tabular}
  \vspace{-10pt}
 \end{table}
}

Next, we evaluated the index merging algorithm with 100, 200,  280, and 500 hospitals. 
Each hospital is assigned with 10 patients and each patient has 20 SNPs. 
The results  are shown in Table~\ref{tab:igdthree}. 
Notably, we observed that the time cost of \emph{IndexMerge} algorithm increases quadratically with the increasing number of hospitals. 
In addition, we also evaluated the fast approach for \emph{IndexMerge} (introduced in Section~\ref{sec:indim}). 
Our results show that the fast approach is more than 290 times faster than the above method, while the search time over the merged index increases around 6 percent with the fast approach.

As discussed, the benefit of applying the \emph{IndexMerge} algorithm is to reduce the search time (i.e., to search over a merged index of multiple hospitals rather than searching over separate indices of individual hospitals). 
To justify this, we analyzed and compared the time costs of the \emph{Search} and \emph{SearchOverMergeIndex} algorithms.
Table~\ref{tab:igdthree} shows that the time cost of \emph{Search} algorithm increases linearly with the increasing number of hospitals while the time cost of \emph{SearchOverMergedIndex} algorithm increases sub-linearly.
When the number of hospitals reaches 100, \emph{SearchOverMergedIndex} algorithm has significant advantage compared to the \emph{Search} algorithm.


 
 {
 \footnotesize
 \begin{table}
 \vspace{-10pt}
 \centering
  \caption{Performance of index merging algorithm (in terms of time cost in seconds) with different number of hospitals. Each hospital has 10 patients and each patient has 20 SNPs.}
   \label{tab:igdthree}
  \label{tab:pcsnh}
 \begin{tabular}{|l|l|l|l|l|l|}
 \hline 
Number of Hospitals & 100 & 200 & 280 & 500 \\
 \hline 
IndexMerge (s) & 2363 & 9059  & 18895  & 59982  \\
 \hline 
SearchOverMergedIndex (s) & 3.9   &  7.68    & 10.7  &  18.4\\ 
 \hline
Search (s)  &  5.8 &   10.8  & 15.2 & 27.2  \\
 \hline 
 \end{tabular}
  \vspace{-10pt}
 \end{table}
}

We also compared the proposed mechanism with the state-of-the-art, including Wang et al.'s~\cite{wang2015efficient}, Asharov et al.'s~\cite{asharov2017privacy}, and Thomas et al.'s \cite{schneider2019episode} schemes. We did the comparison on a single hospital's dataset (as compliant with the settings in~\cite{wang2015efficient} and~\cite{asharov2017privacy}), in which the number of patients is 2850, the number of SNPs per patient is 3350221 (the whole sequence), and the query includes all 3350221 SNPs. 
We first implemented Wang et al.'s scheme~\cite{wang2015efficient} including (i) \emph{protocol one}, which includes a bucketing technique to improve the secure computation of set difference size and (ii) \emph{protocol two}, which replaces the square operation of \emph{protocol one} with an estimation of normal distribution. 
We show the results in Table~\ref{tab:w2015}. We observed that our proposed scheme performs almost $63$ times faster than the best case of \emph{protocol two} of Wang et al.'s scheme. Moreover, our scheme provides higher accuracy than~\cite{wang2015efficient}.
 Next, we compared the proposed scheme with Asharov et al.'s ~\cite{asharov2017privacy} and and Thomas et al.'s~\cite{schneider2019episode} schemes using the performance numbers reported in their papers. We show the comparison result in Table~\ref{tab:ash2017}.
We observed that the running times of Asharov et al.'s and Thomas et al's schemes are sensitive to the number of used SNPs in the query. The run-times of~\cite{asharov2017privacy} and~\cite{schneider2019episode} increase linearly with the number of SNPs in the query while the run-time of our proposed scheme remains constant.
For example, when the number of SNPs in the query reaches to 3350221, the query times of both Asharov et al.'s  and Thomas et al.'s schemes exceed 1100 seconds, which is more than $95$ times slower than our proposed scheme.

{
\footnotesize
\begin{table}
\centering
\caption{Comparison of the proposed mechanism with Wang et al.'s scheme~\cite{wang2015efficient}. $k$ and $l$ are the number of iterations and adopted hash functions respectively.}
\label{tab:w2015}
\begin{tabular}{|c|c|c|c|}
\hline
  \multicolumn{4}{|c|}{Protocol one in~\cite{wang2015efficient}}   \\
\hline 
$k$ & $l$ & Run-time (s)
&   Accuracy \\
\hline
3 & 256 
& 1697 
&  90\% \\ 
\hline 
5 & 256 
& 2683 
& 96\% \\
\hline
 \multicolumn{4}{|c|}{Protocol two in~\cite{wang2015efficient}}   \\
\hline 
$k$ & $l$ & Run-time (s)
&  Accuracy \\
\hline
3 
& 256 
& 730 
&  93\% \\ 
\hline 
5 & 256 
& 1060 
& 96\% \\
\hline
 \multicolumn{4}{|c|}{Proposed mechanism}   \\
\hline 
\multicolumn{2}{|c|}{Bloom filter} & &   \\ 
capacity& error rate& Run-time (s) 
&Accuracy \\
\hline 
$2^{21}$  & 0.01 & 11.59 
&    $\geq 99$ \% \\
\hline 
\end{tabular}
\vspace{-10pt}
\end{table}
 }

{
\footnotesize
\begin{table}
    \centering
    \caption{Comparison of the proposed scheme with Asharov et al.'s ~\cite{asharov2017privacy} and Thomas et al.'s \cite{schneider2019episode} schemes. $\perp$ means the accuracy is not reported.}
    \begin{tabular}{|c|c|c|c|}
        \hline 
         \# of  & \# of SNPs per patient & Run-time (s) & Accuracy \\
         patients & (\# of SNPs in the query) & & \\
         \hline 
         \multicolumn{4}{|c|}{Asharov et al.'s scheme}   \\
         \hline 
         100 & 714  & 0.26 & 94.28\% \\
         \hline
         100 & 1950 & 0.68 & 99.67\% \\
         \hline 
         \multicolumn{4}{|c|}{Thomas et al.'s scheme}   \\
         \hline 
         1000 & 1000 & 1.2 & $\perp$ \\
         \hline 
         1000 & 75M & 24480 & $\perp$ \\
         \hline 
         \multicolumn{4}{|c|}{Proposed mechanism}   \\
         \hline 
         2850 & 20 & 10.036 &   $\geq$ 99\%
         \\
         \hline
         2850 & 2000 & 11  &  $\geq$ 99\% \\
         \hline 
         2850 & 3350221 & 11.59 &  $\geq 99$ \% \\
         \hline 
    \end{tabular}
    \vspace{-10pt}
    \label{tab:ash2017}
\end{table}
 }

Next, we evaluated the performance of ASI sharing focusing on a single ASI with 20, 200, and 2000 associated SNPs, respectively.  We first assumed that the index search result only contains one pseudonym and each patient record only contains one ASI. 
The results are shown in the Table~\ref{tab:asione}. 
We observed that the time costs of \emph{Encrypt}, \emph{SharedKeyGen}, \emph{TokenGen}, \emph{TokenAdjust}, and \emph{ASISearch} algorithms increase linearly with the increasing number of associated SNPs with the ASI. On the other hand, the time cost of \emph{ASIDecrypt} algorithm is almost constant with increasing number of SNPs. 

 {
 \footnotesize
\begin{table}
\centering
\caption{Performance (in terms of time cost in seconds) of ASI sharing mechanism with a single ASI that is associated with different number of SNPs.}
\label{tab:asione}
\begin{tabular}{|l|c|c|c|}
\hline 
Number of SNPs   & 20 & 200 & 2000  \\
\hline 
ASIEncrypt (s)  &   2.95    &   28.87   &  293.4   \\
\hline 
SharedKeyGen (s)   & 0.092& 0.86  & 8.55 \\
\hline 
TokenGen (s)   & 0.083 & 0.83 & 8.3  \\
\hline 
TokenAdjust (s)  & 0.487 & 4.86 & 48.6 \\
\hline 
ASISearch (s)  & 2.3 & 24.7 & 244.3 \\
\hline 
ASIDecrypt (ms)   & 0.066 & 0.07 &0.068 \\         
\hline
\end{tabular}
 \vspace{-10pt}
\end{table}
 }

Finally, we considered a more general case in which each patient has different number of ASIs, each having 20 associated SNPs. The result are shown in Table~\ref{tab:asitwo}. 
We observed that with increasing number of ASIs, the time costs of \emph{SharedKeyGen},  \emph{TokenGen},  \emph{TokenAdjust}, and \emph{ASIDecrypt} algorithms are almost constant. Also, the time costs of \emph{Encrypt} and \emph{ASISearch} algorithms increase linearly with increasing number of ASIs. 

 {
 \footnotesize
\begin{table}
\centering
\caption{Performance (in terms of time cost in seconds) of ASI sharing scheme for different number of ASIs. Each ASI has 20 associated SNPs.}
\label{tab:asitwo}
\begin{tabular}{|l|c|c|c|c|}
\hline 
Number of ASIs &  10 & 100 & 1000&7388  \\
\hline 
ASIEncrypt (s) &   29.055&   290.007   &  2885.7  & 21298.1   \\
\hline 
SharedKeyGen (s) & 0.094 & 0.093& 0.097 & 0.11  \\
\hline 
TokenGen (s) & 0.084 & 0.084 & 0.084 & 0.084  \\
\hline 
TokenAdjust (s)& 0.487& 0.493 & 0.494 &0.494    \\
\hline 
ASISearch (s)& 2.506 & 3.74 & 14.39 & 89.3 \\
\hline
ASIDecrypt (ms) & 0.065 & 0.074 &0.076& 0.073 \\
\hline
\end{tabular}
\vspace{-10pt}
\end{table}
}

\vspace{-5pt}
\subsection{Experiments on Amazon EC2}
To show the scalability of our scheme, we also conducted experiments on Amazon EC2~\cite{awsec2} and especially evaluated the \emph{QueryGen}, \emph{Search}, \emph{IndexMerge}, and \emph{SearchOverMergedIndex} algorithms.  

\vspace{-5pt}
\subsubsection{Data Model}
Using genomic data of 2850 patients (from OpenSNP dataset~\cite{OpenSNP}), we extracted the statistics of the observed SNPs. Using these, we synthetically generated 10000 patients, which is consistent with the previous work, e.g., 
\cite{wang2015efficient}, \cite{schneider2019episode}. 
In detail, we first assigned SNP IDs to 10000 patient records based on the extracted distribution. Then, following the extracted statistics, we assigned a SNP value for each SNP of each patient record. 
We then assigned the generated patients to 100 hospitals (each hospital has 100 patient records). 
In the following experiments, we first built the index for each hospital. The index is built based on all the SNPs (the maximum is 3350221) of each patient record.

\vspace{-5pt}
\subsubsection{Results}

\begin{figure*}[t]
     \centering
     \begin{subfigure}[t]{0.3\textwidth}
         \centering
         \includegraphics[width=\textwidth]{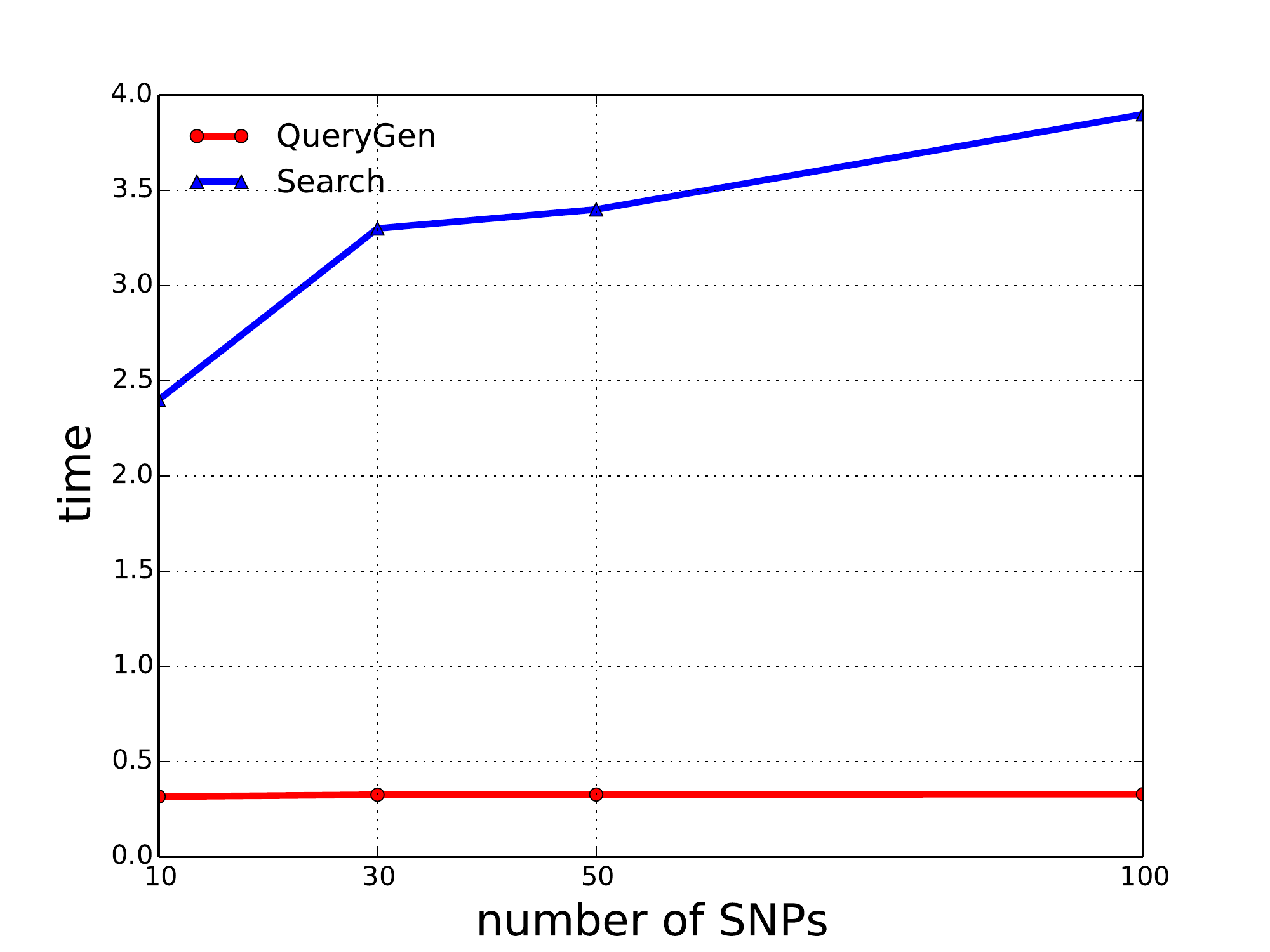}
         \caption{Performance (in terms of time cost in seconds) of \emph{QueryGen} and \emph{Search} with different number of SNPs in the query.}
         \label{fig:awsqgsearch}
     \end{subfigure}
     \hfill
     \begin{subfigure}[t]{0.3\textwidth}
         \centering
         \includegraphics[width=\textwidth]{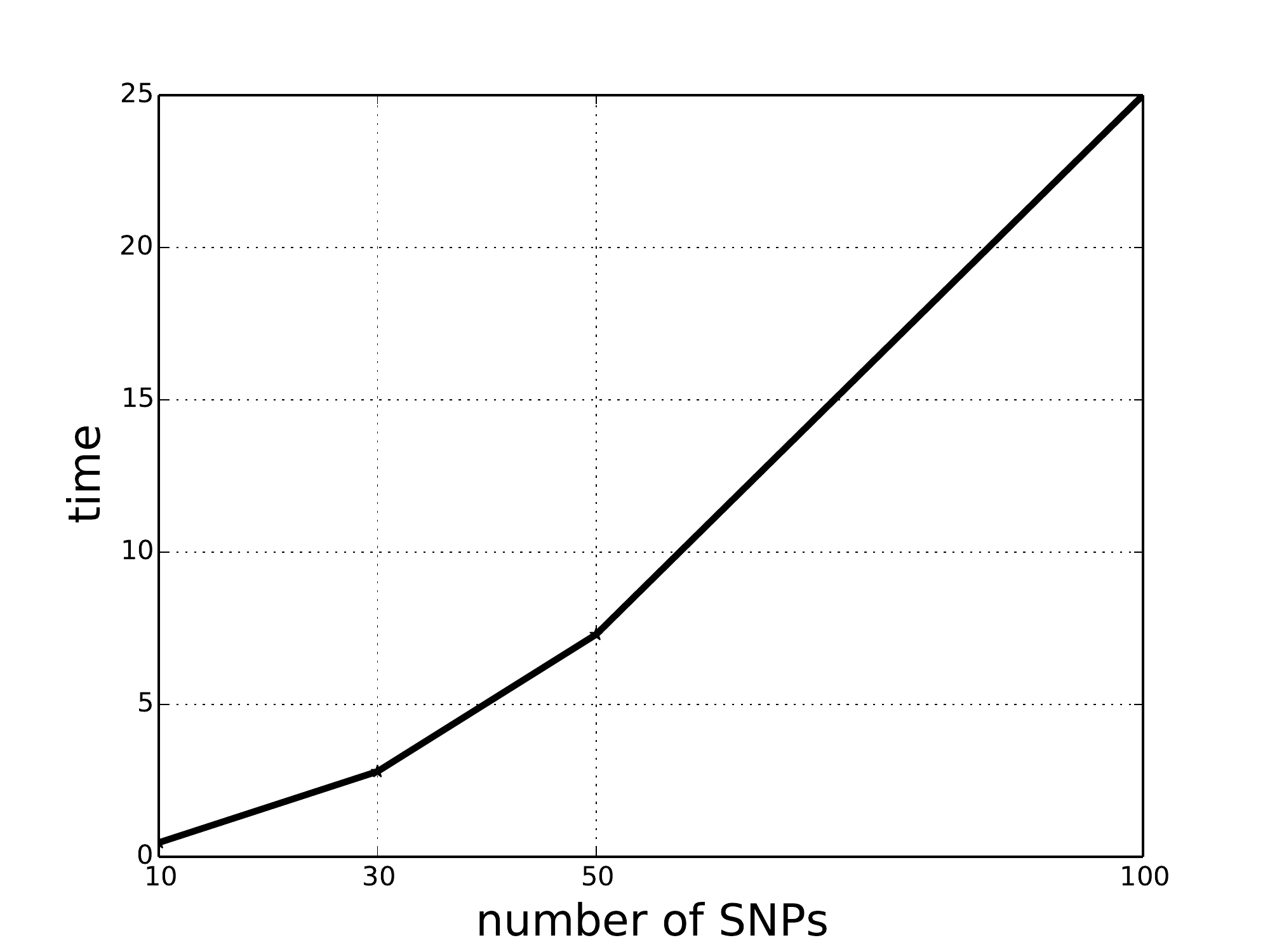}
         \caption{Performance (in terms of time cost in seconds) of \emph{IndexMerge} with different number of merged indices.}
         \label{fig:awsindexmerge}
     \end{subfigure}
     \hfill
     \begin{subfigure}[t]{0.34\textwidth}
         \centering
         \includegraphics[width=\textwidth]{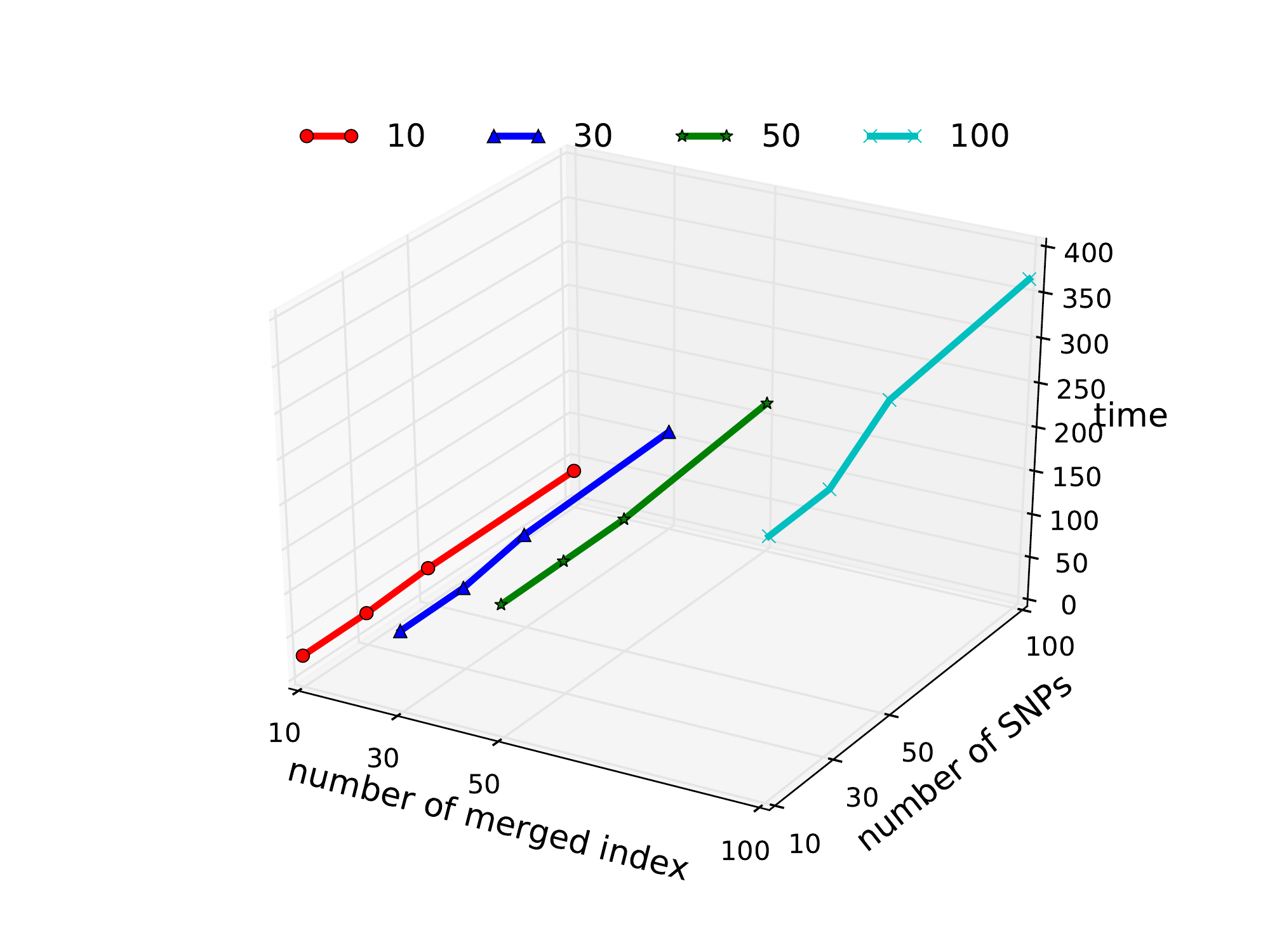}
         \caption{Performance (time cost in seconds) of \emph{SearchOverMergedIndex} with different number of SNPs in the query. Red, blue, green, and cyan represents merged indices that are formed using 10, 30, 50, and 100 indices, respectively.}
         \label{fig:awssomergeindc}
     \end{subfigure}
        \caption{Experimental result from Amazon EC2 platform.}
        \label{fig:awssomergeind}
        \vspace{-14pt}
\end{figure*}
First, we evaluated the performance of the 
\emph{IndexGen} and \emph{HierarchialIndexGen} algorithms. 
We observed that the time costs for \emph{IndexGen} is 742.51 seconds and \emph{HierarchicalIndexGen} is 896.93 seconds for building a hierarchical index.
In Figure~\ref{fig:awsqgsearch}, we show the performance of the \emph{QueryGen} algorithm for different number of SNPs in the query. We observed that the time cost of \emph{QueryGen} increases slightly with increasing number of SNPs in the query. 
We then evaluated the time cost of the \emph{Search} algorithm and observed (in Figure~\ref{fig:awsqgsearch}) that search time increases linearly with increasing number of SNPs in the query.

Next, in Figure~\ref{fig:awsindexmerge}, we show the performance of the \emph{IndexMerge} algorithm. 
For the evaluation of the \emph{IndexMerge} algorithm, we adopted the fast approach described in Section~\ref{sec:indim}. 
We observed that the time cost of the \emph{IndexMerge} algorithm increases superlinearly when increasing the number of indices. Finally, in Figure~\ref{fig:awssomergeindc}, we show the time cost of the \emph{SearchOverMergedIndex} algorithm while varying the number of merged indices and SNPs in the query.  
We observed that the number of SNPs has a limited impact on the search efficiency compared to the number of indices. Moreover, we observed that when the number of indices reaches 50, the time cost of sequentially calling \emph{Search} is at least 170 seconds, while the maximum time cost of 
\emph{SearchOverMergedIndex} is 166 seconds. 
This also supports the experimental result obtained on the local machine, which shows that 100 indices are enough to benefit from index merging.


\vspace{-5pt}
\section{Discussion}
\label{sec:discuss}
\vspace{-5pt}

Here, we discuss the evaluation results and potential applications.

\vspace{-5pt}
\subsection{Evaluation Results and Practicality}
\vspace{-5pt}

In general, we observe that the experimental results (in Section~\ref{sec:perf}) on a single machine and Amazon EC2 are consistent. In addition, we observe that the search process benefits from merging process when the number of indices  (e.g., hospitals) exceeds a certain number (50 in our large-scale experiments). Based on the results of the evaluation, we show that the proposed scheme is scalable with respect to increasing number of hospitals, number of patients per hospital, number of SNPs per patient, and number of ASIs per patient. We also show the superiority of the proposed scheme with respect to the state-of-the-art in terms of its practicality. In particular, we show that the proposed scheme has significant advantage for scenarios that include large number of SNPs per patient and large number of SNPs in the query.

The most time consuming part of the proposed scheme is the \emph{IndexMerge} algorithm, which is conducted by the CSP and executed infrequently. We showed that a fast alternative of the \emph{IndexMerge} significantly reduces the run-time while it slightly increases the time needed for the search operation.
Also, the implementation of \emph{IndexMerge} can be further optimized (especially for the fast approach) via concurrent programming.



\vspace{-5pt}
\subsection{Alternative Usecases}
\vspace{-5pt}

The proposed scheme can also be used in existing health-related online social networks. For instance, the well-known online health information exchange platform \emph{PatientsLikeMe}~\cite{patientslikeme} (PLM) already attracts more than 0.6 million members including physicians, researchers, and patients. Currently PLM includes only the ASI (phenotype, ethnicity, disease conditions, treatment, etc.) of its members. However, it is not very far fetched to assume that such a platform also starts storing genomic data of its members in the near future. PLM is a popular platform even among physicians to learn about the treatment procedures of other physicians. Using our proposed scheme, genomes of individuals can be kept encrypted at the PLM. PLM can index such genomes and let the physicians conduct privacy-preserving similar patient tests on them. As a result of the test, the PLM can connect the physicians of the corresponding patients so they can exchange information about their patients. Similarly, using the proposed scheme, and assuming a cloud-based genomics service provider (such as Google Genomics) act as the CSP, research labs and hospitals that keep their data on the cloud can query each other's databases in a privacy-preserving way.

\vspace{-5pt}
\section{Conclusion}\label{sec:conclusion}
\vspace{-5pt}

In this paper, we have proposed a privacy-preserving and efficient solution for the similar patient search problem among several hospitals. To achieve this, we have proposed a novel privacy-preserving index structure. To improve the efficiency of the search operation, we have developed a hierarchical index structure (to index each hospital's dataset with low memory requirement) and a novel privacy-preserving index merging mechanism that generates a common search index from individual indices of each hospital. We have also considered the search for medical information (e.g., diagnosis and treatment) that is associated with genomic data of a patient. 
We have developed a scheme that allows access to this information via a fine-grained access control policy. Via simulations on real and synthetic genomic data, we have shown the practicality and efficiency of the proposed scheme. We believe that the proposed scheme will further facilitate the use of genomic data in clinical settings and pave the way for personalized medicine. 
In future work, we will focus on supporting dynamic datasets and we will extend our scheme to support batch search.

\bibliographystyle{acm}
\bibliography{genome}

\begin{appendices}

\section{Background on Cryptographic Tools}
\label{app:background}

Here, we provide background on common cryptographic tools such as asymmetric bilinear groups and Bloom filter.

\subsection{Asymmetric Bilinear Groups}
Let $G_1$ and $G_2$ be two distinct groups of prime order $p$ and $g_1\in G_1$ and $g_2 \in G_2$ be the generators of $G_1$ and $G_2$, respectively. Let $e: G_1 \times G_2 \rightarrow G_T$ be a function which maps two elements from $G_1$ and $G_2$ to a target group $G_T$  of prime order $p$. The tuple $(G_1, G_2, G_T, p, e)$ is an asymmetric bilinear group if following properties hold:  \\
(a) the group operations in $G_1$, $G_2$, $G_T$ can be computed efficiently.  \\
(b) $e$ can be computed efficiently.   \\
(c) $e$ is non-degenerate: $e(g_1, g_2) \neq 1$.   \\
(d) $e$ is bilinear: for all $a,b \in \mathbb{Z}_p$, $e(g_1^a, g_2^b)=e(g_1,g_2)^{ab}$.

\subsection{Bloom Filter}
A Bloom filter is a bit array used to efficiently check the existence of an element in a set~\cite{bloom1970space}. At the beginning, all the values of the array elements are set to $0$. There exists a family of $k$ different hash functions, each function mapping a data item to a position inside the array. Consequently, each data item is represented by $k$ non-zero bits inside the bit array. Even if a data item has not been mapped to the bit array, there is still a probability for the $k$ corresponding bits to be non-zero because of the other data items represented in the array. Such a situation is called ``false positive''. Let $m$ be the length of the bit array and $n$ be the number of distinct data items mapped to the array. The false positive probability is expressed as $(1-e^{\frac{-(kn)}{m}})^k$, and it gets the smallest value when $k=\ln2 \frac{m}{n}$. 

\section{System's Core Procedures}
\label{sec:scsap}
We provide the core procedures of the proposed scheme in Table~\ref{tab:csiss}. 

\begin{table*}[h!]
\caption{System's Core Procedures. 
The first column includes the names of the APIs. 
The inputs and outputs of the APIs are shown in columns 2 and 3. 
The last column represents the corresponding module of the API in Figures~\ref{fig:initialisation}, \ref{fig:clientauthorization}, and \ref{fig:queryprocessing}. 
The value of the last column consists of two parts. The first part (before dot) represents the number of the corresponding figure and the second part represents the number of the module in the figure.}
\label{tab:csiss}
\centering
\begin{tabular}{|c|c|c|c|}
\hline
\textbf{API Call}  & \textbf{Input} & \textbf{Output} & \textbf{Steps in Figure~\ref{fig:initialisation}, \ref{fig:clientauthorization} and \ref{fig:queryprocessing}}\\
\hline
Setup & $\lambda$ & initial functions and parameters & \ref{fig:initialisation}.1 \\
\hline 
IndexGen & $K$, $\textit{Dict}_i^\textit{BF}$, $\textit{PK}_{i,1}$, $\textit{SK}_{i}$ & $\Delta_i$, $r_i$, $\sigma_i$ & \ref{fig:initialisation}.2 \\
\hline 
SNPEncrypt & $ K_{\alpha_i}$, $\mathbf{S}_{i, \textit{SNP}}$ & $C_{i,\textit{SNP}}$ & \ref{fig:initialisation}.3 \\
\hline 
ASIEncrypt & $K_i$, $K_{\beta_i}$, $\textit{Dict}_i^\textit{ASI}$ & $\textit{Dict}_i^\textit{C}$ & \ref{fig:initialisation}.4\\
\hline
HierarchicalIndexGen  & $\Delta_i$ & $\Delta_i^H$ & \ref{fig:initialisation}.5\\
\hline
IndexMerge & $I_\textit{ind}$ & $\varrho$ & \ref{fig:initialisation}.6\\
\hline 
\hline
SharedKeyGen & $K_i$, $K_c$, $\mathbf{S}_{i,SNP}$ & $\delta_i$, \textit{$cbf_i$} &   \ref{fig:clientauthorization}\\
\hline 
\hline 
QueryGen & $K$, $S_c$, $\varepsilon_c$, $k_c$,$\sigma_i$ & $Q$ & \ref{fig:queryprocessing}.1\\
\hline
TokenGen & $K_c$, $S_{c,SNP}$ & \textit{TK} & \ref{fig:queryprocessing}.2\\
\hline 
Search & $\Delta_i^H$, $r_i$, $Q$ & \textit{Dict} & \ref{fig:queryprocessing}.3\\
\hline
SearchOverMergedIndex & $\varrho$, $Q$ & \textit{Dict} & \ref{fig:queryprocessing}.3\\
\hline 
TokenAdjust & $TK$, $\delta_i$, $\textit{cbf}_i$ & $\textit{TK}^*$ & \ref{fig:queryprocessing}.4\\
\hline 
ASISearch & $\textit{TK}^*$, $\textit{Dict}_i^\textit{C}$, $\eth$, $\textit{MK}_i$ & $C_{C_1}$ & \ref{fig:queryprocessing}.5\\
\hline
ASIDecrypt & $S_{\beta_i}$, $C$ & $S_\textit{ASI}$ & \ref{fig:queryprocessing}.6\\
\hline
\end{tabular}
\end{table*}

\section{Key parameters and functions}
\label{sec:keyp_func}

 We list the frequently used parameters and functions in Table~\ref{tab:ipf}.
\begin{table}[h!]
    \centering
    \caption{Key parameters and functions.}
    \label{tab:ipf}
    \begin{tabular}{|l|l|}
    \hline
        $G_1$/$G_2$/$G_T$ &  a group of prime order $p$ \\
    \hline
        $e$ & a bilinear mapping from $G_1$, $G_2$ to $G_T$ \\
    \hline
        \textit{bf} & maps the input into a Bloom filter  \\
    \hline 
        \textit{cbf} & maps the input into a 
        customized Bloom filter \\
    \hline     
    $\textit{Dict}_i^\textit{BF}$ & a dictionary of hospital $i$ that stores pairs including 
                  \\& (i) pseudonym of a 
                 patient and
                (ii) Bloom filter     \\& 
                output of the corresponding
                patient's genome \\
    \hline
     $\textit{Dict}_i^\textit{CBF}$ & a dictionary of hospital $i$ that stores \{pseudonym, \\& customized Bloom filter\} pairs \\
    \hline 
    $\textit{Dict}_M^\textit{CBF}$ & a merged dictionary from multiple hospitals \\
    \hline
    $\textit{Dict}_i^\textit{ASI}$ & a dictionary of hospital $i$ that stores \{pseudonym, \\& ASI plaintext\} pairs \\
    \hline
    $\textit{Dict}_i^\textit{C}$ & a dictionary of hospital $i$ that stores \{pseudonym, 
     \\& ASI ciphertext\} pairs \\
    \hline
    $\mathbf{S}$ & a set of SNPs \\ 
    \hline 
    $\mathbf{S}_{i, \textit{SNP}}$ & a set of SNPs from hospital $i$ \\
    \hline 
    $\mathbf{S}_{i,\textit{SNP}}^{\textit{ASI}_i}$ & a set of SNPs related to $\textit{ASI}_i$ from hospital $i$\\
    \hline 
        $H_0$ & maps two strings to an random string \\
    \hline
        $H_1$ & maps a string to an element of group $G_1$ \\
    \hline
        $H_2$ & maps two elements from $G_T$ to a string  \\
    \hline
        $F$ & a pseudorandom function (PRF) \\
    \hline 
        $K$ & a secret key,
                   used by all hospitals 
                  and \\
                  & approved clients \\
    \hline 
        $K_c$ & a secret key selected by a client \\
    \hline 
       $\varepsilon_c$ & a threshold of minimum
       number of matching \\& SNPs set  by client $c$ \\
    \hline 
     $k_c$ & a threshold of maximum retrieved 
     \\& result set by client $c$ \\
    \hline 
    
        $K_i$ & a  secret key selected by hospital $i$ for ASI  
       \\&  encryption   and  shared key  generation \\
              
    \hline 
        $K_{\alpha_i}$ & a symmetric encryption key of hospital 
        $i$ for SNP \\& encryption \\
    \hline  
        $K_{\beta_i}$ & a symmetric encryption key of
                  hospital $i$ for ASI  \\
                  & encryption, which is
                  shared with authorized clients  \\
    \hline 
    $K_{\gamma_i}$ & a symmetric encryption key of  
     hospital $i$ randomly \\& selected from $G_T$ \\
    \hline
    $\textit{PK}_{i,1}$, $\textit{SK}_{i}$ & a pair of public/private keys  
    selected by \\& hospital $i$ for  signature \\
    \hline 
    $\textit{PK}_{i,2}$, $\textit{MK}_i$ & a pair of public and master keys of CPABE \\& selected by   hospital $i$ and shared with the CSP \\
    \hline 
    \end{tabular}
\end{table}

\section{Details of the Algorithms in the Initialisation Phase in Section~$5.2$}
\label{app:initialization}

Here, we provide the details of the algorithms introduced in the initialization phase of the proposed scheme (in Section~\ref{sec:indini}). We show the details of the index generation, ASI encryption, hierarchical clustering, and index merging in Algorithms~\ref{alg:indgen}, \ref{alg:asiencrypt}, \ref{alg:hirarchicalindex}, and \ref{alg:indexmerge}, respectively. 


\alglanguage{pseudocode}
\begin{algorithm}
\caption{\emph{IndexGen}}
\label{alg:indgen}
\begin{algorithmic}[1]
\Require a secret key $K$ of PRF, dictionary $Dict_i^\textit{BF}$, public key $\textit{PK}_{i,1}$ of signature, private key $\textit{SK}_{i,1}$ of signature 
\Ensure a pair $\Delta_i$ of dictionary and public key, random string ~$r_i$, and signature~$\sigma_i$
\State initialize a dictionary $Dict_i^{CBF}$
\State $r_i \xleftarrow{\textit{R}}\{0,1\}^\lambda$
\ForAll{$\textit{ID}_i \in Dict_i^\textit{BF}$}
\State \text{initialize a customized Bloom filter}~$\textit{cbf}_{ID_i}$
\State $\textit{bf}_{\textit{ID}_i} \gets$ \textit{$Dict_i^\textit{BF}$}[$\textit{ID}_i$]
 \ForAll{\textit{integer~pos} $\in$ [1,$|\textit{bf}_{\textit{ID}_i}|$]~\text{s. t.}~ \textit{$\textit{bf}_{\textit{ID}_i}$}[\textit{pos}]=1} 
  \State $\zeta_1 \gets F(K, \textit{pos})$
  \State $\zeta_2 \gets F(r_i, \zeta_1)$
  \State $ \textit{cbf}_{ID_i}.\textit{add}(\zeta_2)$
\EndFor
 \State $Dict_i^\textit{CBF}[\textit{ID}_i] \gets \textit{cbf}_{ID_i}$
\EndFor
\State $ hr \gets H_0(\textit{PK}_{i,1}, r_i)$
\State $\sigma_i \gets Sign(\textit{SK}_{i}, hr)$
\State $\Delta_i=(Dict_i^\textit{CBF}, \textit{PK}_{i,1}, r_i)$
\State \Return $\Delta_i, \sigma_i$
\end{algorithmic}
\end{algorithm}



\alglanguage{pseudocode}
\begin{algorithm}
\caption{\emph{ASIEncrypt}}
\label{alg:asiencrypt}
\begin{algorithmic}[1]
\Require a secret key~$K_i$ of hospital $i$, a secret key~$K_{\beta_i}$ of symmetric encryption,  dictionary~$Dict_i^\textit{ASI}$ storing  ASI plaintext
\Ensure dictionary $Dict_i^\textit{C}$ storing ASI ciphertext  
\State initialize a dictionary $Dict_i^{C}$
\ForAll{$\textit{ID}_i \in Dict_i^\textit{ASI}$}
\State $C_{ID_i} \gets \phi$ 
\ForAll{ $(\textit{ASI}, S_{\textit{ID}_i,\textit{SNP}}^\textit{ASI}) \in  Dict_i^\textit{ASI}[\textit{ID}_i]$}
\State $\tau \xleftarrow{\textit{R}}\{0,1\}^\lambda$
\State $ \mathbf{\theta}  \gets \phi$
\ForAll{ $ (\textit{SNP.ID}, \textit{SNP.val}) \in \mathbf{S}_{\textit{ID}_i, \textit{SNP}}^\textit{ASI}$}
\State $v \leftarrow \textit{SNP.ID}\circ SNP.val$
\State $h \gets H_2(\tau, e(H_1(v), g_2)^{1/K_i})$
\State $ \mathbf{\theta}  \gets  \mathbf{\theta}  \cup h$
\EndFor
\State $C_1 \gets \textit{AES.Enc}(K_{\beta_i}, \textit{ASI})$
\State $K_{\gamma_i} \xleftarrow{R} G_T$
\State $C_2 \gets \textit{AES.Enc}(K_{\gamma_i}, C_1)$
\State $\mathcal{A} \gets \textit{CPABE.AP}(\theta)$
\State $C_3 \gets \textit{CPABE.Enc}(\textit{PK}_{i,2}, K_{\gamma_i}, \mathcal{A})$
\State $\mathbf{C}_{ID_i} \gets \mathbf{C}_{ID_i} \cup (C_3, C_2, \tau)$
\EndFor
\State $Dict_i^\textit{C}[\textit{ID}_i] \gets  \mathbf{C}_{ID_i}$ 
\EndFor
\State \Return $Dict_i^\textit{C}$
\end{algorithmic}
\end{algorithm}

\alglanguage{pseudocode}
\begin{algorithm}
\caption{\emph{HierarchicalIndexGen}}
\label{alg:hirarchicalindex}
\begin{algorithmic}[1]
\Require \text{unclustered index}~$\Delta_i$
\Ensure \text{hierarchical index}~$\Delta_i^{H}$
\State $(Dict_i^\textit{CBF}, \textit{PK}_{i,1}, r_i) \gets \Delta_i$
\State $(\textit{Tr}_i, Dict_i^{H}) \gets \textit{HC(Dict}_i^\textit{CBF})$
\State $\Delta_i^{H} \gets (\textit{Tr}_i, Dict_i^{H}, \textit{PK}_{i,1}, r_i)$
\State \Return $\Delta_i^{H}$
\end{algorithmic}
\end{algorithm}


\alglanguage{pseudocode}
\begin{algorithm}
\caption{\emph{IndexMerge}}
\label{alg:indexmerge}
\begin{algorithmic}[1]
\Require a set of indices~$\mathbf{I}_{ind}$
\Ensure a merged index~$\varrho$
\State initialize a dictionary $Dict_{temp}$
\State $\mathbf{\Upsilon} \gets \phi$
\ForAll{$ind \in \mathbf{I}_{ind}$}
\State $(\textit{Tr}_i, Dict_i^\textit{CBF}, \textit{PK}_{i,1}, r_i) \gets ind$
\State $\mathbf{\Upsilon} \gets \mathbf{\Upsilon} \cup (\textit{PK}_{i,1},r_i)$
\ForAll{leaf \emph{node} $\in$ $\textit{Tr}_i$}
\State $Dict_{temp}[\textit{node.ID} \circ r_i] \gets Dict_i^\textit{CBF}[\textit{node.ID}]$
\EndFor
\EndFor
\State $(\textit{Tr}_{M}, Dict_{M}^\textit{CBF}) \gets \textit{HC(Dict}_{temp})$
\State $\varrho \gets (\mathbf{\Upsilon}, \textit{Tr}_{M}, Dict_{M}^\textit{CBF})$
\State \Return $\varrho$
\end{algorithmic}
\end{algorithm}

\section{Illustration of ASI Encryption Process}
\label{sec:asienca}
\begin{figure}
\centering
\includegraphics[scale=0.8]{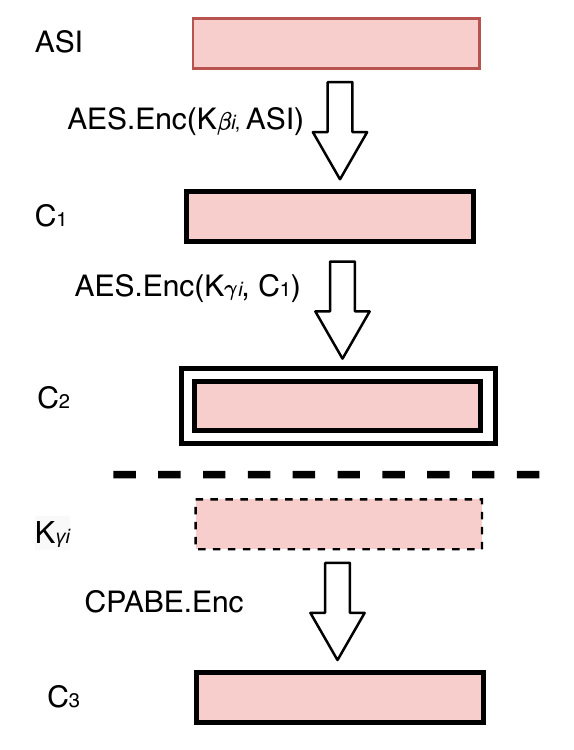}
\caption{Overview of ASI encryption. \emph{ASI} is first encrypted using AES encryption and secret key $K_{\beta_i}$ to obtain $C_1$. Then, $C_1$ is encrypted using AES and with secret key $K_{\gamma_i}$ to obtain $C_2$. Finally,  $K_{\gamma_i}$ is encrypted to $C_3$ by using CPABE.}
\label{fig:asienc}
\end{figure}

We discuss the ASI encryption process in Section~\ref{sec:data_enc}. We show an illustration of ASI encryption in Figure~\ref{fig:asienc}.

\section{Details of the Algorithms in the Client Authorization Phase in Section~$5.3$}
\label{app:client_auth}
Here, we provide the details of the algorithm (in Section \ref{sec:cau}) to generate a shared key in the client authorization phase of the proposed scheme. The details to generate a shared key is shown in Algorithm~\ref{alg:sharedkeygen}.


\alglanguage{pseudocode}
\begin{algorithm}
\caption{\emph{SharedKeyGen}}
\label{alg:sharedkeygen}
\begin{algorithmic}[1]
\Require a secret key~$K_i$ of hospital $i$, a secret key $K_c$ of a client,a set~$\mathbf{S}_{i,SNP}$ of authorized SNPs
\Ensure shared key~($\delta_i$, $cbf_i$) 
\State initialize a customized Bloom filter $cbf_i$
\State  $\delta_i \gets g_2^{K_c/K_i}$
\ForAll{$(SNP.ID, SNP.val) \in \mathbf{S}_{i,SNP}$}
\State $v \leftarrow SNP.ID \circ SNP.val$
\State $t_v \gets H_1(v)^{1/K_c}$
\State $cbf_i$.add($t_v$)
\EndFor
\State \Return ($\delta_i$, $cbf_i$)
\end{algorithmic}
\end{algorithm}

\section{Details of the Algorithms in the Query Processing Phase in Section~$5.4$}
\label{app:query_process}
Here, we provide the details of the algorithms introduced in the query processing phase of the proposed scheme (in Section~\ref{sec:query_processing}). We show the details of query generation, token generation, search conducted on a single index, and the \emph{Insert} function in Algorithms~\ref{alg:querygen}, \ref{alg:tokengen}, \ref{alg:searchsingle}, and \ref{alg:insert}, respectively. We also show the detailed steps of search over a merged index, \emph{InsertConditionally} function, and token adjustment in Algorithms~\ref{alg:searchmerge}, \ref{alg:secondinsert}, and \ref{alg:tokenadjust}, respectively. Finally, we show the details of ASI search and ASI decryption in Algorithms~\ref{alg:asisearch} and \ref{alg:asidecrypt}.

\alglanguage{pseudocode}
\begin{algorithm}
\caption{\emph{QueryGen}}
\label{alg:querygen}
\begin{algorithmic}[1]
\Require a secret key~$K$ of PRF, a set $\mathbf{S}_c$ of SNPs, \text{threshold}~$\varepsilon_c$, top~$k_c$, and signature~$\sigma_i$
\Ensure query $Q$
\State \text{initialize a Bloom filter} $bf$
\ForAll{$(SNP.ID, SNP.val) \in \mathbf{S}_{c,SNP}$}
        \State  $bf.add(SNP.ID \circ SNP.val)$
\EndFor
    \State $\mathbf{E} \gets \phi$ 
    \State $count \gets 0$
    \ForAll{integer $pos \in [1, |bf|]$~\textit{such that}~$bf[pos]=1$}
        \State $\zeta_1 \gets F(K, pos)$
        \State count $\gets$ count+1
        \State $\mathbf{E} \gets \mathbf{E} \cup \zeta_1$
    \EndFor
\State $\varepsilon_c$ $\gets$ $\frac{count}{|S|}\cdot \varepsilon_c$
    \State \Return Q $\gets$ ($\mathbf{E}$, $\varepsilon_c$, $k_c$, $\sigma_i$)
\end{algorithmic}
\end{algorithm}


\alglanguage{pseudocode}
\begin{algorithm}
\caption{\emph{TokenGen}}
\label{alg:tokengen}
\begin{algorithmic}[1]
\Require  a secret key~$K_c$, a set $\mathbf{S}_{c, SNP}$ of SNPs 
\Ensure \text{token} $\mathbf{TK}$ 
\State $\mathbf{TK} \gets \phi$
\ForAll{$(SNP.ID, SNP.val) \in \mathbf{S}_{c,SNP}$}
\State $v \leftarrow SNP.ID \circ SNP.val$
\State $tk \gets H_1(v)^{1/K_c}$
\State $\mathbf{TK} \gets \mathbf{TK} \cup tk$
\EndFor
\State \Return $\mathbf{TK}$
\end{algorithmic}
\end{algorithm}


\alglanguage{pseudocode}
\begin{algorithm}
\caption{\emph{Search}}
\label{alg:searchsingle}
\begin{algorithmic}[1]
\Require \text{hierarchical index}~$\Delta_i^{H}$, \text{random string}~$r_i$, and \text{query}~$Q=(\mathbf{E},\varepsilon_c, k_c,\sigma_i)$
\Ensure a dictionary $Dict$ with $k_c$ entries
\State $(Tr_i, Dict_i^{H}, PK_{i,1}) \gets \Delta_i^{H}$ 
\State  $Dict=\phi$
\If{$verfiy(PK_{i,1}, \sigma_i, H_0(PK_{i,1}, r_i))=False)$}
    \State \Return None
\EndIf
\State initialize a queue $qu$ and a customized Bloom filter~$cbf_c$
\ForAll{$\zeta_1 \in \mathbf{E}$}
\State $\zeta_2 \gets F(r_i, \zeta_1)$
\State $cbf_c.add(\zeta_2)$
\EndFor
\State $root_i \gets Tr_i $
\State $qu.push(root_i)$
\While{$qu$ is not empty}
\State \emph{n} $\gets qu.pop()$
\State $sim \gets \frac{Dict_i^{H}[n.ID_i] \cdot cbf_c}{|Dict_i^{H}[n.ID_i]|\cdot|cbf_c|}$
\Comment{cosine similarity}
\If{$sim \geq \varepsilon_c$}
    \If{\emph{n} is a leaf}
        \State $Dict \gets Insert(Dict, n.ID_i, sim)$
    \Else \If{\emph{n} has a left child}
            \State \emph{leftchild} $\gets$ the left child of the \emph{n}
            \State $qu$.push(\emph{leftchild})
         \EndIf
        \If{\emph{n} has a right child}
             \State \emph{rightchild} $\gets$ the right child of the \emph{n}
             \State $qu$.push(\emph{rightchild})
        \EndIf
    \EndIf
\EndIf
\EndWhile
\State \Return $Dict$
\end{algorithmic}
\end{algorithm}


\alglanguage{pseudocode}
\begin{algorithm}
\caption{\emph{Insert}}
\label{alg:insert}
\begin{algorithmic}[1]
\Require dictionary~$Dict$, pseudonym~\textit{$ID_i$}, similarity score~$sim$ 
\Ensure dictionary~$Dict$
\If{len($Dict$)<$k_c$}
\State $Dict(ID_i) \gets sim$
\State  \Return $Dict$
\EndIf
\State find the pair ($ID_{temp}$, $sim_{temp}$) such that
\State $sim_{temp}=min_{(ID_{temp}, sim_{temp}) \in Dict}$ 
\If{$sim_{temp}< sim$}
\State delete the pair ($ID_{temp}$, $Dict_k[ID_{temp}]$) from $Dict$
\State $Dict[ID_i] \gets sim$
\EndIf
\State \Return $Dict$
\end{algorithmic}
\end{algorithm}


\alglanguage{pseudocode}
\begin{algorithm}
\caption{\emph{SearchOverMergedIndex}}
\label{alg:searchmerge}
\begin{algorithmic}[1]
\Require Merged index~$\varrho$, query~$Q=(\mathbf{E}, \sigma_{s}, \varepsilon_c, k_c)$
\Ensure dictionary~$Dict$ with $k_c$ entries
\State $(\mathbf{\Upsilon}, Tr_M, Dict_M^{CBF})  \gets \varrho$
\State initialize dictionary $Dict$ with $k_c$ entries 
\State initialize a customized Bloom filter $cbf_c$ 
\State $\mathbf{\Re} \leftarrow \phi$
\ForAll{ $(PK_{i,1}, r_i) \in \mathbf{\Upsilon}$}
\ForAll{ $\sigma_i \in \mathbf{\sigma}_{s}$}
\If{$verify(PK_{i,1}, \sigma_i, H(PK_{i,1},r_i)$)=False}
\State $\mathbf{\Re} \leftarrow \mathbf{\Re}\cup r_i$  
\ForAll{$\zeta_1 \in \mathbf{E}$}
\State $\zeta_2 \leftarrow F(r_i, \zeta_1)$
\State $cbf_c.add(\zeta_2)$
\EndFor
\EndIf
\EndFor
\EndFor
\If{$cbf_c \neq \phi$}
\State $root \gets Tr_M $
\State initialize a queue $qu$
\State $qu.push(root)$
\While{$qu$ is not empty}
\State \emph{n} $\gets qu.pop()$
\State $sim \gets \frac{Dict_M^{CBF}[n.ID_i]\cdot cbf_c}{|Dict_M^{CBF}[n.ID_i]|\cdot |cbf_c|}$
\Comment{cosine similarity}
\If{$sim \geq \varepsilon_c$}
    \If{\emph{n} is a leaf}
        \State Dict $\gets$ InsertConditionally(Dict, $\theta_r$, $n.ID_i$, sim)
    \Else \If{\emph{n} has a left child}
            \State \emph{leftchild} $\gets$ the left child of the \emph{n} 
            \State $qu$.push(\emph{leftchild})
         \EndIf
        \If{\emph{n} has a right child}
             \State \emph{rightchild} $\gets$ the right child of the \emph{n}
             \State $qu$.push(\emph{rightchild})
        \EndIf
    \EndIf
\EndIf
\EndWhile
\EndIf
\State \Return Dict
\end{algorithmic}
\end{algorithm}



\alglanguage{pseudocode}
\begin{algorithm}
\caption{\emph{InsertConditionally}}
\label{alg:secondinsert}
\begin{algorithmic}[1]
\Require dictionary~$Dict$, pseudonym~$ID_i$, random string set~$\mathbf{\Re}_r$, similarity score~\textit{sim}
\Ensure dictionary $\textit{Dict}$
\State \textcolor{red}{$(ID_i^*, r_i^*) \gets ID_i$}
\If{\textcolor{red}{$r_i^* \in \mathbf{\Re}$ } }
\State $Dict \gets Insert(Dict, ID_i^*,sim)$
\EndIf
\State \Return Dict
\end{algorithmic}
\end{algorithm}


\alglanguage{pseudocode}
\begin{algorithm}
\caption{\emph{TokenAdjust}}
\label{alg:tokenadjust}
\begin{algorithmic}[1]
\Require a set~$\mathbf{TK}$ of \text{raw tokens}, a \text{shared key}~($\delta_i$,$cbf_i$) 
\Ensure a set~$\mathbf{TK}^*$ of \text{valid tokens}
\State $\mathbf{TK}^* \gets \phi$
\ForAll{$tk \in \mathbf{TK}$ such that $cbf_i(tk)=True$}
\State $tk^* \gets e(tk, \delta_i)$
\State $\mathbf{TK}^* \gets \mathbf{TK}^* \cup tk^* $
\EndFor
\State \Return $\mathbf{TK}^*$
\end{algorithmic}
\end{algorithm}


\alglanguage{pseudocode}
\begin{algorithm}
\caption{\emph{ASISearch}}
\label{alg:asisearch}
\begin{algorithmic}[1]
\Require  valid token~$\mathbf{TK}^*$, encrypted ASI data~$Dict_i^{C}$, a set~$\mathbf{\eth}$ of pseudonyms,  a master key~$MK_i$ of CPABE 
\Ensure  a set $\mathbf{C_{C_1}}$ of ASI ciphertext
\State $\mathbf{C}_{C_1} \gets \phi$
\ForAll{$ID_i \in \mathbf{\eth}$}
\ForAll{$(C_3,C_2, \tau) \gets Dict_i^{C}[ID_i]$}
\State $\vartheta \gets \phi $
\ForAll{$tk \in \mathbf{TK}^*$}
\State $h \gets H_2(\tau, tk)$
\State $\mathbf{\vartheta} \gets \mathbf{\vartheta} \cup h$
\EndFor
\State $sk \gets CPABE.KeyGen(MK_i, \vartheta)$
\If{$sk \neq \phi$ }
\State $K_{\gamma_i} \gets CPABE.Decrypt(sk, C_3)$
\If{$K_{\gamma_i} \neq False$}
\State $C_1 \gets AES.Decrypt(K_{\gamma_i}, C_2)$
\State $\mathbf{C}_{C_1} \gets \mathbf{C}_{C_1} \cup C_1$
\EndIf
\EndIf
\EndFor
\EndFor
\State \Return $C_{C_1}$
\end{algorithmic}
\end{algorithm}

\alglanguage{pseudocode}
\begin{algorithm}
\caption{\emph{ASIDecrypt}}
\label{alg:asidecrypt}
\begin{algorithmic}[1]
\Require a secret key $ K_{\beta_i}$ of symmetric encryption, a set~$\mathbf{C}_{C_1}$ of ASI ciphertext
\Ensure a set~$\mathbf{S}_{ASI}$ of plaintext of ASIs
\State $\mathbf{S}_{ASI} \gets \phi$
\ForAll{$c \in \mathbf{C}$}
\State $ASI \gets AES.Decrypt( K_{\beta_i}, c)$
\State $\mathbf{S}_{ASI} \gets \mathbf{S}_{ASI} \cup ASI $
\EndFor
\State \Return $\mathbf{S}_{ASI}$
\end{algorithmic}
\end{algorithm}

\section{Leakage Function}
\label{sec:leak_func}
The leakage function plays an important role in the privacy analysis as it defines the information that is allowed to be acquired by the adversary. Following previous work~\cite{chase2010structured, curtmola2011searchable,zhu2019privacy}, we consider the following as the allowed leaked information: (i) the size pattern, (ii) search pattern, and (iii) access pattern. 
The size pattern includes the size of encrypted query, encrypted genomic sequences, encrypted index, and encrypted ASI. 
The search pattern represents the relationship between a query and the retrieved result. 
The access pattern represents the access path to certain data records. 
To define the leakage function formally, we first present the formal definitions of the aforementioned patterns. 

\noindent\textbf{Size pattern ($\mu$):} Let $C_\textit{SNP}$=\{$C_{1,\textit{SNP}}$, $\cdots$, $C_{n,\textit{SNP}}$\},
$C_\textit{ASI}$=\{$C_{1,\textit{ASI}}$,$\cdots$, $C_{n, \textit{ASI}}$\}, and 
$C_\textit{ind}$=\{$(\Delta_1, \sigma_1), \cdots, (\Delta_m, \sigma_m)$ \} 
be the ciphertexts of genome sequences, ASI, and index stored at the CSP, respectively (where \textit{n} is the total number of records stored in the CSP and \textit{m} is the total number of indices). Also, let \textit{Q} and \textit{TK} be the client's query for the index and token of ASI, respectively. The size pattern is defined as $\mu$=\{$|C_\textit{SNP}|$, $|C_\textit{ASI}|$, $|C_\textit{ind}|$, $|Q|$, $|\textit{TK}|$\}. 

\noindent\textbf{Search pattern ($\nu$):}  Let $\{q_1, \cdots, q_p\}$ be $p$ consecutive queries and \{$S_{1,\textit{ASI}} \cdots, S_{q,\textit{ASI}} $\} be the corresponding retrieved ASIs. 
Then, $\nu$ is defined as a two dimensional matrix and $\nu[i, j]=1$ if a retrieved ASI exists in the $j$th position of the patient $i$'s record.  

\noindent\textbf{Access pattern ($\xi$):} Let $C_\textit{ind}$ be the set of encrypted indices and $C_\textit{ASI}$ be the set of encrypted ASIs at the CSP. Let \{$S_{1,\textit{ASI}} \cdots, S_{q,\textit{ASI}} $\} be ASIs retrieved by queries \{$q_1=(Q_1, \textit{TK}_1), \cdots, q_p=(Q_p, \textit{TK}_p)$\}. Then, the access pattern is defined as $\xi$=\{$(C_{ind}(Q_1)$, $C_\textit{ASI}(C_{ind}(Q_1)$, $\textit{TK}_1$), $S_{1,\textit{ASI}}$), $\cdots$, $(C_{ind}(Q_p)$, $C_\textit{ASI}(C_{ind}(Q_p)$, $\textit{TK}_p$), $S_{p,\textit{ASI}}$) \}.

The leakage function captures the leakage of the above defined patterns and it is defined as follows. 

\noindent\textbf{Leakage function ($\mathcal{L}$):} Let $C_\textit{SNP}$, $C_{ind}$, $C_\textit{ASI}$, and $q=(Q, \textit{TK})$ be the encrypted genome sequences, encrypted index, encrypted ASIs, and a query, respectively. The leakage function is defined as $\mathcal{L}=\{C_\textit{SNP}, C_{ind},C_\textit{ASI}, Q, \textit{TK},  \mu, \nu, \xi \}$. The output of $\mathcal{L}$ is the revealed bits that are not supposed to be disclosed. 

The leakage function $\mathcal{L}$ is used to control the information leakage of allowed requests (e.g., query request).  

\section{Privacy of Genomic Data}\label{sec:genomic_privacy}
We formulate the privacy of genomic data as a game between a challenger and an adversary, 
which includes both privacy of index and genome sequence. 
First, the adversary selects two datasets $DB_{0,SNP}$ and $DB_{1, SNP}$ of SNPs. 
Each item of dataset $DB_{i, SNP}~(i \in \{0,1\})$ includes two components: patient pseudonym and corresponding SNPs of the patient. 
Then, the adversary sends the two databases to the challenger. The adversary is allowed to send adaptive queries with constraint on the information leakage before making the final decision about which dataset is utilized by the challenger. 

Let $\Pi_{SNP}$=\{\emph{Setup}, \emph{IndexGen}, \emph{QueryGen}, \emph{IndexMerge}, \emph{Search},\\ \emph{SearchOverMergedIndex}\} be a set of algorithms of the proposed scheme that are related to genomic data privacy. 
For a probabilistic polynomial time (PPT) adversary $\mathit{Adv}$, 
the advantage function $ADV^{\Pi_{SNP}}_{\mathit{Adv}}$ is defined as follows. $ADV^{\Pi_{SNP}}_{\mathit{Adv}}$ =$Pr(b^*=b) - \frac{1}{2}$, where $b$ and $b^*$ are defined in the following game which 
evaluates the probability of breaking the proposed scheme. We describe the key steps of the game between the challenger and the adversary below.

\noindent\textbf{Init:} The adversary $\mathit{Adv}$ submits two datasets $DB_{0, SNP}$ and $DB_{1, SNP}$ to the challenger with the same number of records and index structure. 

\noindent\textbf{Setup:} The challenger generates the initial functions, parameters, and keys. Details can be found in Section~\ref{sec:indini}. 

\noindent\textbf{Phase 1:} The adversary is allowed to obtain the ciphertexts of genome, index, and query by adapatively submitting ciphertext requests, index requests, and query requests to the challenger.
\\
\noindent\emph{Ciphertext request}: The adversary selects a dataset of genome sequence and submits it to the challenger to request its ciphertext. The selected dataset is not limited to $DB_{0, SNP}$ and $DB_{1, SNP}$.  
\\
\noindent\emph{Index request}: The adversary selects a dataset that is different from $DB_{0,SNP}$ and $DB_{1, SNP}$ and submits it to the challenger to request its index. 
 
\noindent\textbf{Challenge:} The challenger randomly selects a bit $b$ from $\{0, 1\}$ and encrypts $DB_{b, SNP}$ to generate $C_{b, SNP}$. It also generates encrypted index $C_{ind_b}$ and sends $C_{b, SNP}$ and $C_{ind_b}$ to the adversary. 


\noindent\textbf{Phase 2:} The adversary adaptively submits query request $Q$ in addition to the ciphertext and index requests described in the Phase 1. 
\\
\noindent\emph{Query request}: The adversary selects target SNPs, sets two thresholds, attaches the signature of target index, and sends them to the challenger for asking the query with constraint $\mathcal{L}(S_{0, SNP}, Q, C_{ind_0})=\mathcal{L}(S_{1, SNP}, Q, C_{ind_1})$.

\noindent\textbf{Guess:} The adversary $\mathcal{A}$ outputs $b^*$ as a guess of $b$.

The privacy of the genomic data is preserved against selective chosen plaintext attack if in the above scheme $\Pi_{SNP}$ the adversary $\mathit{Adv}$ has negligible advantage. Explicitly, the advantage function $ADV^{\Pi_{SNP}}_{\mathit{Adv}}$ should be a negligible function in parameter $\lambda$.

\section{Privacy of ASI}
\label{app:ASI_privacy_game}
Similar to privacy of genomic data, we also define the privacy of ASI via a game between a challenger and the adversary. In a nutshell, the adversary sends two ASIs to a challenger and the adversary sends adaptive queries with constraint on the leakage function before providing its guess of the applied ASI. The formal definition of this game is below.

Informally, the ASI privacy is defined as a game in which the adversary sends two ASIs to a challenger and the adversary sends adaptive queries with constraint on the leakage function before providing its guess of the applied ASI. The formal definition is defined as follows. 

Let $\Pi_{ASI}$=\{\emph{Setup}, \emph{ASIEncrypt}, \emph{SharedKeyGen}, \emph{TokenGen}, \emph{TokenAdjust}, \emph{ASISearch}, \emph{ASIDecrypt}\} be a set of algorithms of the proposed scheme that are related to ASIs. For a PPT adversary $\mathit{Adv}$, the advantage of wining the game is defined as 
$ADV_{\mathit{Adv}}^{\Pi_{ASI}}=Pr(b^*=b)-\frac{1}{2}$, where $b^*$ and b are defined in the following game. 
\\
\textbf{Init:} The adversary selects $ASI_0$ and $ASI_1$ with same size and number of associated SNPs and submits them to the challenger. \\
\textbf{Setup:} The challenger runs \emph{Setup} to start the system with initial functions, parameters, and keys. Details can be found in Section~\ref{sec:indini}. 
\\
\textbf{Phase 1:} The adversary adaptively submits requests in one of the following types:
 \\
\emph{Ciphertext request:} The adversary submits an ASI associated to a set of specified SNPs and requests for its ciphertext. The selected ASI is not limited to the previously uploaded two ASIs. 
\\
\emph{Token request:} The adversary submits a set of SNPs and requests the corresponding search token.  
\\
\emph{Shared key request:} The adversary submits a set of SNPs and requests for a shared key with constraint that the shared key can decrypt either both $ASI_0$ and $ASI_1$ or none of them. 
\\
\textbf{Challenge:} The challenger randomly selects a bit $b$ from $\{0, 1\}$ and encrypts $ASI_b$ by calling function \emph{ASIEncrypt} before sending the result to the adversary. 
\\
\textbf{Phase 2:} The adversary repeats Phase 1.
\\
\textbf{Guess:} The adversary outputs its guess $b^*$ for $b$. 

We claim the scheme $\Pi_{ASI}$ is privacy-preserving against selective chosen plaintext attack if the PPT adversary $\mathit{Adv}$ has negligible advantage in winning the above game.

\section{Proof of Theorem~$6.2$}
\label{app:thorem_indtwo}

 \begin{proof}
We prove Theorem~\ref{thr:indtwo} by assuming if there exists an adversary $\mathit{Adv}$ that can distinguish the two views of $DB_{0, SNP}$ and $DB_{1, SNP}$, then there exists a simulator $\mathcal{B}$ that can break either the semantic security of AES encryption or the randomness of PRF.
In the following, we follow the previously defined game (in Appendix~\ref{sec:genomic_privacy}) with a simulator $\mathcal{B}$ trying to break the proposed scheme.

\noindent\textbf{Init:} A simulator $\mathcal{B}$ selects and submits two datasets of genome sequences $DB^*_{0,SNP}$ and $DB^*_{1, SNP}$ to the adversary $\mathit{Adv}$ and challenger with same number of records and index structure. \\
\textbf{Setup:} The challenger runs \emph{Setup} to set initial parameters and functions. 
\\
\textbf{Phase 1}: The adversary adaptively submits one of the following requests to the simulator $\mathcal{B}$.
\\
\emph{Ciphertexts request}:  $\mathcal{B}$ directly submits the dataset from adversary  to the challenger and sends the ciphertext from the challenger to the adversary.
The dataset is not limited to $DB^*_{0,SNP}$ and $DB^*_{1, SNP}$. 
\\
\emph{Index request}: The simulator $\mathcal{B}$ directly sends the submitted dataset to the challenger and sends the returned index from the challenger back to the adversary $\mathcal{A}$.
The submitted dataset is not among  $DB^*_{0,SNP}$ and $DB^*_{1, SNP}$.
\\
\textbf{Challenge}: The challenger randomly selects a bit $b$ and generates the ciphertext and index of $DB^*_{b,SNP}$ by invoking \emph{SNPEncrypt} and \emph{IndexGen}, respectively. The challenger sends the newly generated ciphertext and index to the simulator $\mathcal{B}$.
The simulator $\mathcal{B}$ sends them to the adversary.  
\\
\textbf{Phase 2:} The adversary submits the following query request to simulator $\mathcal{B}$ in addition to repeating the Phase 1.   
\\
\emph{Query request}: Simulator $\mathcal{B}$ uploads the submitted query request to the challenger and sends the result back to the adversary.  
\\
\textbf{Guess:} The adversary outputs its guess $b^*$ to the simulator $\mathcal{B}$ and the simulator $\mathcal{B}$ outputs the same guess.

According to the initial assumption, the adversary $\mathit{Adv}$ has significant advantage in breaking the defined experiment in Appendix~\ref{sec:genomic_privacy}. In the proof, the described experiment strictly follows the defined experiment, and hence the simulator $\mathcal{B}$ has significant advantage in guessing the correct answer. 
Thus, the simulator $\mathcal{B}$ can distinguish the ciphertext $C^*_{0,SNP}$ and index $C^*_{ind_0}$ from $C^*_{1,SNP}$ and index $C^*_{ind_1}$ with significant advantage in the experiment.
Since $C^*_{b,SNP}$ is the ciphertext obtained from AES encryption and index $C^*_{ind_b}$ is the output of a PRF, the simulator $\mathcal{B}$ successfully breaks one of them if the simulator $\mathcal{B}$ cannot learn significant information from the ciphertext request, index request, and query request.

First, we analyse the ciphertext request. Since each ciphertext is generated as a result of AES encryption, the security is guaranteed by the robustness of AES. 
According to the assumption, the adopted AES achieves semantic security and the simulator $\mathcal{B}$ cannot learn significant information from the ciphertext. 
Second, the index request is only applicable to the datasets that are different from the challenged datasets. 
Moreover, each index is randomized by a random string, and hence the simulator $\mathcal{B}$ cannot correlate any two different indices. 
Third, the query request is constrained by the leakage function. The request will be released if and only if the request causes the same leakage of two challenged datasets. As the leakage is the same, the simulator $\mathcal{B}$ cannot learn significant information to distinguish the challenged ciphertext and index. 
Based on the above analysis,  we can conclude that in the above experiment, the simulator $\mathcal{B}$ should break either the AES encryption or the randomness of PRF.  

\end{proof}

\end{appendices}
\end{document}